\documentclass{pasj00}

\usepackage{ccaption}
\usepackage{multirow}

\draft

\begin{document}
\SetRunningHead{M. Fujishita et al.}{Molecular Loop in the Galactic Center}
\Received{2009/1/2}
\Accepted{}

\title{Discovery of Molecular Loop 3 in the Galactic Center:\\
       Evidence for a Positive-Velocity Magnetically Floated Loop \\
       towards $L=355^{\circ}$--$359^{\circ}$}

\author{Motosuji \textsc{Fujishita},\altaffilmark{1} 
        Kazufumi \textsc{Torii},\altaffilmark{1}
        Natsuko \textsc{Kudo},\altaffilmark{1}
        Tokuichi \textsc{Kawase},\altaffilmark{1} \\
        Hiroaki \textsc{Yamamoto},\altaffilmark{1}
        Akiko \textsc{Kawamura},\altaffilmark{1}
        Norikazu \textsc{Mizuno},\altaffilmark{2}
        Toshikazu \textsc{Onishi},\altaffilmark{3} \\
        Akira \textsc{Mizuno},\altaffilmark{4} 
        Mami \textsc{Machida},\altaffilmark{1}
        Kunio \textsc{Takahashi},\altaffilmark{5}
        Satoshi \textsc{Nozawa},\altaffilmark{6} \\
        Ryoji \textsc{Matsumoto},\altaffilmark{7}
        and 
        Yasuo \textsc{Fukui}\altaffilmark{1}
        }

\altaffiltext{1}{Department of Astrophysics, Nagoya University, Chikusa-ku, Nagoya 464-8602}
\altaffiltext{2}{National Astronomical Observatory of Japan, Osawa, Mitaka, Tokyo 181-8588}
\altaffiltext{3}{Department of Physical Science, Osaka prefecture University, Sakai, Osaka 599-8531}
\altaffiltext{4}{Solar-Terrestrial Environment Laboratory, Nagoya University, Chikusa-ku, Nagoya 464-8601}
\altaffiltext{5}{Japan Agency for Marine-Earth Science and Technology, Kanazawa-ku, Yokohama, Kanagawa 236-0001, Japan}
\altaffiltext{6}{Department of Science, Ibaraki University, 2-1-1 Bunkyo, Mito, Ibaraki 310-8512}
\altaffiltext{7}{Faculty of Science, Chiba University, Inage-ku, Chiba 263-8522}

\email{motosuji@a.phys.nagoya-u.ac.jp, fukui@a.phys.nagoya-u.ac.jp}


%


\maketitle

\begin{abstract}
 We have discovered a molecular dome-like feature towards $355^{\circ} \leq l \leq 359^{\circ}$ and $0^{\circ} \leq b \leq 2^{\circ}$. 
 The large velocity dispersions of 50--100 km s$^{-1}$ of this feature are 
 much larger than those in the Galactic disk and indicate that 
 the feature is located in the Galactic center, probably within $\sim1$ kpc of Sgr A$^{*}$. 
 The distribution has a projected length of $\sim600$ pc and height of $\sim300$ pc from the Galactic disk 
 and shows a large-scale monotonic velocity gradient of $\sim130$ km s$^{-1}$ per $\sim600$ pc. 
 The feature is also associated with H\emissiontype{I} gas having a more continuous spatial and velocity distribution 
 than that of \atom{C}{}{12}\atom{O}{}{}. We interpret the feature as a magnetically floated loop similar to loops 1 and 2 
 and name it "loop 3". 
 Loop 3 is similar to loops 1 and 2 in its height and length but is different from loops 1 and 2 
 in that the inner part of loop 3 is filled with molecular emission. 
 We have identified two foot points at the both ends of loop 3. 
 H\emissiontype{I}, \atom{C}{}{12}\atom{O}{}{} and \atom{C}{}{13}\atom{O}{}{} datasets 
 were used to estimate the total mass and kinetic energy of loop 3 to be 
 $\sim3.0 \times 10^{6} \Mo$ and $\sim1.7 \times 10^{52}$ ergs. 
 The huge size, velocity dispersions and energy are consistent 
 with the magnetic origin the Parker instability as in case of loops 1 and 2 
 but is difficult to be explained by multiple stellar explosions.  
 We argue that loop 3 is in an earlier evolutionary phase than loops 1 and 2 based 
 on the inner-filled morphology and the relative weakness of the foot points. 
 This discovery indicates that the western part of the nuclear gas disk of $\sim1$ kpc radius 
 is dominated by the three well-developed magnetically floated loops 
 and suggests that the dynamics of the nuclear gas disk is strongly affected by the magnetic instabilities.
\end{abstract}

\section{Introduction}

 Molecular clouds are the sites of star formation and the distribution and dynamics of molecular gas 
 should play a crucial role in the evolution of the Galaxy. 
 Such evolution may occur in different manners between the Galactic center and disk 
 because the stellar gravitational filed is considerably different between them, 
 making the pressure much stronger in the central region than in the disk. 
 It is important to understand the physical implications of the difference 
 in our efforts to elucidate galactic structure and evolution.
 
 \par
 
 Most of the molecular gas in the Galactic center is concentrated within about 300 pc 
 of the Galactic center and is called the "Central Molecular Zone" (hereafter CMZ, \cite{mor1996}) 
 including the Sgr A and B2 molecular clouds (e.g., \cite{sco1975}; \cite{fuk1977}; \cite{gus1983}; \cite{oka1998b}; 
 \cite{tsu1999}; \cite{mar2004}; see also for a recent review, \cite{gus2004}). 
 There are also molecular features with weaker intensities outside the CMZ up to nearly 1 kpc from the center. 
 They include Clump 1 at $(l,\ b) \sim (355^{\circ},\ 0^{\circ})$, Clump 2 at $(l,\ b) \sim (3^{\circ},\ 0^{\circ})$ 
 and the 5-deg feature at $(l,\ b) \sim (5^{\circ},\ 0^{\circ})$ \citep{ban1977} and at least several weak CO emission features 
 detected at a lower resolution of 8\farcm8 (\cite{dam1987}; \cite{bit1997}). 
 These features outside the CMZ received little attention so far.
 
 \par
 
 Most recently, \citet{fuk2006} discovered molecular features, loops 1 and 2, 
 having negative velocities
 in $355^{\circ} \leq l \leq 358^{\circ} $ and $0^{\circ} \leq b \leq 2^{\circ} $ 
 based on an analysis of the NANTEN Galactic plane survey CO dataset. 
 These loops have heights of $\sim2$ degrees from the galactic plane and the projected lengths of $\sim3$--4 degrees. 
 The loops have two foot points that are brightest spots in the CO emission on each end 
 at galactic latitudes around $\sim0.8$ degrees. 
 The velocity dispersions of the molecular gas are $\sim50$ km~s$^{-1}$ in the foot points; 
 such dispersions are characteristic to the molecular gas in the Galactic center 
 but is much larger than those of the disk molecular clouds 
 whose velocity spans are less than $\sim10$ km~s$^{-1}$. 
 \citet{fuk2006} present discussion that the loops cannot be due to supershells, 
 because the velocity distribution does not fit what is expected for an expanding shell. 
 Instead, the two loops are interpreted by magnetic flotation caused by the Parker instability (\cite{pak1966}), 
 where the foot points are a natural outcome of the flotation; 
 the gas falls down along the loops to form shock fronts with enhanced intensity and large velocity dispersion.
 Theoretical studies of magneto-hydrodynamics (MHD) indeed show that a magnetic flotation loop 
 with two foot points is a general phenomenon in a galaxy (\cite{mat1988}; \cite{fuk2006}). 
 It is now an issue of keen interest if there are more magnetically floated features 
 in the rest of the Galactic center and if the magnetic instability is a mechanism 
 which dominates the gas dynamics in the Galaxy. 
 
 \par
 
 In the CMZ, the molecular gas shows high temperatures and violent motions. 
 The temperature is derived to be $\sim50$--$300$ K using CO, NH$_{3}$, H$_{2}$, H$_{3}^{+}$ lines 
 (\cite{mar2004}; \cite{hut1993}; \cite{rod2001}; \cite{oka2005}). 
 The velocity dispersion is 15--50 km s$^{-1}$, which is significantly larger than 
 that in the molecular clouds in the Galactic disk (e.g., \cite{mor1996}; \cite{gus2004}). 
 The causes of the high temperatures and large velocity dispersions have been the longstanding puzzles 
 in the last few decades. The magnetic field is as strong as 0.1--1 mG 
 as observed and/or suggested by Radio Arc and other non-thermal filaments 
 which are distributed within 1 degree of the center, 
 often vertically to the Galactic plane (\cite{yus1984}; 2004) and 
 by the infrared double helix \citep{mor2006}. 
 The magnetic flotation model has a potential to offer a coherent explanation 
 on the origin of the high temperature and large velocity dispersion of molecular gas 
 in the Galactic center as argued by \citet{fuk2006}.
 
 \par
 
 Subsequent to the discovery of loops 1 and 2 in a negative velocity range 
 with respect to the local standard of rest (LSR) towards 
 $355^{\circ} \leq l \leq 358^{\circ}$ and $0^{\circ} \leq b \leq 2^{\circ}$, 
 we have searched for other loop-like molecular features with the NANTEN Galactic plane survey 
 (GPS) \atom{C}{}{12}\atom{O}{}{} ($J=1$--$0$) dataset and discovered another loop towards 
 $355^{\circ} \leq l \leq 359^{\circ}$, $0^{\circ} \leq b \leq 2^{\circ}$ (hereafter loop 3) 
 in a positive velocity range.  
 We hereafter use the word "velocity" to refer to the velocity with respect to LSR. 
 In this paper, we present results on a newly identified molecular loop 3 
 towards the same field with the loops 1 and 2 based on the NANTEN CO dataset. 
 Section 2 describes the \atom{C}{}{12}\atom{O}{}{} and \atom{C}{}{13}\atom{O}{}{} ($J=$1--$0$) dataset. 
 General properties of magnetic flotation loops are presented both theoretically 
 and observationally in section 3. 
 Section 4 presents results and discussion is given in section 5. 
 Section 6 summarizes the paper. 
 
\newpage

\section{Data set}

\subsection{\atom{C}{}{12}\atom{O}{}{} ($J=1$--$0$) emission}
 The \atom{C}{}{12}\atom{O}{}{} ($J=1$--$0$) dataset towards the Galactic center region was taken with the NANTEN 4-m radio telescope 
 at Las Campanas Observatory in Chile during the period from March 1999 to September 2001. 
 The half-power beamwidth (HPBW) was 2\farcm6 at 115 GHz, the frequency of \atom{C}{}{12}\atom{O}{}{} ($J=1$--$0$). 
 The front end was a 4 K cryogenically cooled Nb superconductor-insulator-superconductor 
 (SIS) mixer receiver \citep{oga1990} that provided a typical system temperature of $\sim280$ K 
 in the single-side band, including the atmosphere towards the zenith. 
 The spectrometer was an acousto-optical spectrometer (AOS) with 2048 channels. 
 The frequency coverage and resolution were 250 MHz and 250 kHz, 
 corresponding to a velocity coverage of 650 km s$^{-1}$ and a velocity resolution of 0.65 km s$^{-1}$, 
 respectively, at 115 GHz. 
 The intensity calibration was made using the chopper-wheel method \citep{kut1981}. 
 The absolute antenna temperature was calibrated by observing $\rho$ Oph East 
 [RA(1950) = \timeform{16h29m20.9s}, Dec(1950) = \timeform{-24D$22'13"}] 
 every 2 hours, whose absolute temperature was assumed to be 15 K. 
 The telescope pointing was measured to be accurate to within $20^{\prime\prime}$ 
 by radio observations of planets in addition to optical observations of stars 
 with a CCD camera attached to the telescope. 
 The observed region was 240 deg$^{2}$ towards $-12^{\circ} \leq l \leq 12^{\circ}$ and $-5^{\circ} \leq b < 5^{\circ}$ 
 at a grid spacing of $4\arcmin$, corresponding to 10 pc at a distance of the Galactic center, 8.5 kpc. 
 In total, 54,000 positions were observed. 
 The integration time per point was 4--9 s, resulting in typical r.m.s.~noise 
 fluctuations of 0.36 K at a velocity resolution of 0.65 km s$^{-1}$.

\subsection{\atom{C}{}{13}\atom{O}{}{} ($J=1$--$0$) emission}
 The \atom{C}{}{13}\atom{O}{}{} ($J=1$--$0$) dataset towards the Galactic center region was taken with 
 the same instrument as the \atom{C}{}{12}\atom{O}{}{} ($J=1$--$0$) observations in October 2003. 
 The half-power beam width (HPBW) was 2\farcm7 at 110 GHz, the frequency of \atom{C}{}{13}\atom{O}{}{} ($J=1$--$0$). 
 The same receiver provided a typical system temperature of $\sim100$ K in the single-side band, 
 including the atmosphere towards the zenith. 
 The same spectrometer provided the frequency coverage and resolution, 250 MHz and 250 kHz, 
 corresponding to a velocity coverage of 620 km s$^{-1}$ and a velocity resolution of 0.62 km s$^{-1}$, 
 respectively, at 110 GHz. 
 The absolute antenna temperature was calibrated by observing $\rho$ Oph East 
 [RA(1950) = \timeform{16h29m20.9s}, Dec(1950) = \timeform{-24D22'13"}] every 2 hours. 
 Absolute radiation temperature of $\rho$ Oph East was assumed to be 10 K. 
 The observed region was 28 deg$^{2}$ towards $-6^{\circ} \leq l < 8^{\circ}$ and $-1^{\circ} \leq b < 1^{\circ}$ 
 at a grid spacing of $2\arcmin$. 
 In total, 25,200 positions were observed. 
 The integration time per point was 10--15 s, resulting in typical r.m.s.~noise 
 fluctuations of 0.20 K at a velocity resolution of 0.62 km s$^{-1}$.

\newpage

\section{General Properties of Magnetic Flotation Loops}

\subsection{Theories}

 Pioneering theoretical studies of galactic magnetic-floatation loops by the Parker instability 
 were made by \citet{mat1988} and \citet{hor1988}. 
 The magnetic flotation is a general consequence of the Parker instability 
 where the gas layer is supported by the stratified magnetic field lines, 
 whose pressure is in balance against the stellar gravity. 
 The two basic parameters in the instability are the pressure scale height $H$ and 
 the Alfv$\mathrm{\acute{e}}$n speed $V_{\mathrm{A}} = B / \sqrt{4 \pi \rho}$, 
 where $B$ is the magnetic field and $\rho$ is the gas density. 
 The height and wavelength of the loop are given by a few times $H$ and $\lambda = 10$ times scale height, 
 respectively. 
 Such magnetically floated loops cause gas flows downward along the magnetic filed lines to the nuclear disk, 
 causing often shock fronts \citep{mat1988}. 
 
 \par
 
 \citet{fuk2006} presented an interpretation that molecular loops 1 and 2 
 are created by the Parker instability in the magnetized nuclear disk 
 and estimate that the field strength is 150 $\mu$G at 400--500 pc 
 from the center by assuming the energy equi-partition between the turbulent gas motion and the magnetic field. 
 These authors made a two-dimensional numerical simulation of loops 1 and 2 and showed 
 that the two loops are successfully reproduced by the magnetic flotation mechanism.
 
 \par
 
 Subsequently, \citet{mac2009} carried out three-dimensional global numerical simulations 
 of the nuclear gas disk and showed that the magneto-rotational instability 
 coupled with the Parker instability works to create more than a few 100 loops 
 over a 1kpc-radius nuclear disk, where Miyamoto-Nagai potential \citep{miy1975} 
 was adopted as the stellar gravitational field. 
 
 \par
 
 The numerical simulation indicates general aspects of the magnetic floatation. 
 The differential rotation in the nuclear disk creates toroidal magnetic field. 
 This initially-azimuthal field configuration tends to have radial components 
 by the magneto-rotational instability which mixes the gas and field at different radii. 
 The molecular gas is dynamically coupled to the magnetic field 
 because the molecular part of the disk is ionized at an ionization degree of 
 $\sim10^{-7}$ by cosmic ray protons (e.g., \cite{gus2004}). 
 Therefore, the gas is floated when a magnetic field is floated as a loop. 
 The gas inside the loop flows down to the disk under the influence of the disk gravity, 
 forming a uniform velocity gradient along the loop. 
 This down flowing gas collides with the disk and often forms shock fronts with enhanced turbulent motions. 
 The gas in the foot point is heated and compressed by the shock fronts. 
 Therefore, the magnetic floatation offers a coherent explanation 
 both on the large velocity dispersions and high temperatures of the nuclear gas disk.
 
 \par
 
 \citet{mac2009} have also shown that one-armed non-axisymmetric density pattern 
 of $m = 1$ mode is developed in the nuclear disk. 
 In the low density region, half of the nuclear disk, 
 magnetic pressure tends to become stronger and prominent magnetic loops are formed preferentially 
 there as compared to the higher density region on the opposite side of the nuclear disk. 
 This fact is consistent with the observational results 
 that about three-fourths of the dense molecular gas, CMZ, 
 is distributed in the positive Galactic longitudes and two loops, loops 1 and 2, 
 are distributed in the negative Galactic longitudes.
 
\subsection{Observational Properties of Loops 1 and 2}

 \citet{fuk2006} analyzed the NANTEN Galactic plane survey dataset 
 and revealed that two molecular loops having negative velocities 
 are located towards $355^{\circ} \leq l \leq 358^{\circ}$ and $0^{\circ} \leq b \leq 2^{\circ}$. 
 These loops have heights of $\sim2$ degrees from the galactic plane 
 and the projected lengths of $\sim3$--4 degrees. 
 The loops have two foot points that are brightest spots in the CO emission on each end 
 at Galactic latitudes around $\sim0.8$ degrees. 
 The velocity dispersions of the molecular gas in the loops are $\sim50$ km s$^{-1}$ 
 at the foot point of loops, 
 which is characteristic to the molecular gas in the Galactic center 
 and much larger than that of the disk molecular clouds whose velocity spans are less than $\sim10$ km s$^{-1}$. 
 Also, the loops show fairly uniform velocity gradients along themselves 
 over a velocity span of $\sim80$ km s$^{-1}$. 
 It is likely that the two loops are located in the Galactic center 
 because the large velocity dispersions are typical to the inner several degrees. 
 These loops are interpreted by magnetic flotation caused by the Parker instability.
 
 \par
 
 Here, we summarize observational signatures of the magnetic floatation loops; 
 (1) Molecular gas is floated from the Galactic plane in a loop-like shape, 
 (2) The loop has two foot points at the both ends which have enhanced velocity dispersions, 
 (3) The floated component which connects the foot points shows a monotonic velocity gradient 
 as a result of the flowing down motion along the loop. 
 We have searched for molecular features which satisfy these observational properties in the NANTEN GPS dataset.
 
\newpage

\section{Results}

\subsection{\atom{C}{}{12}\atom{O}{}{} ($J=1$--$0$) distributions}
 
\begin{figure}
  \begin{center}
    \FigureFile(160mm,150mm){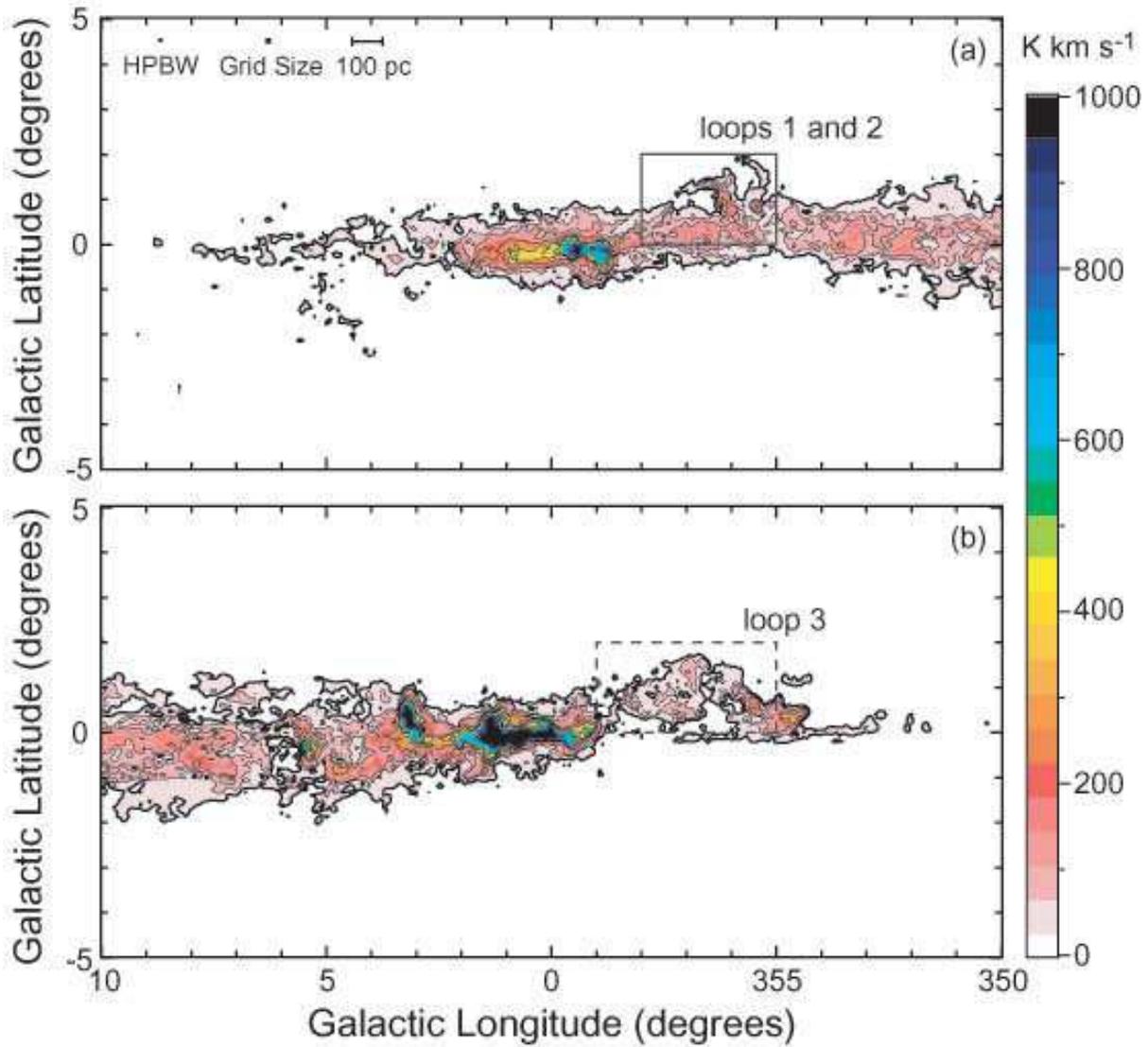}
  \end{center}
  \caption{Velocity-integrated intensity maps of \atom{C}{}{12}\atom{O}{}{} ($J=1$--$0$) emission 
           observed with NANTEN.
           The lowest contour level is 20 K km s$^{-1}$ and denoted by the thick line.
           Contours are plotted at 
           20, 60, 100, 220, 340, 460, 820, 1180, 1540, 1900 K km s$^{-1}$.
           (a) The integrated velocity range is from -300 km s$^{-1}$ to -20 km s$^{-1}$.
           The region of loops 1 and 2 is denoted by the solid line.
           (b) The integrated velocity range is from 20 km s$^{-1}$ to 300 km s$^{-1}$.
           The region of loop 3 is denoted by the broken line.
           (Color Online)}
  \label{ent}
\end{figure}

\begin{figure}
  \begin{center}
    \FigureFile(160mm,125mm){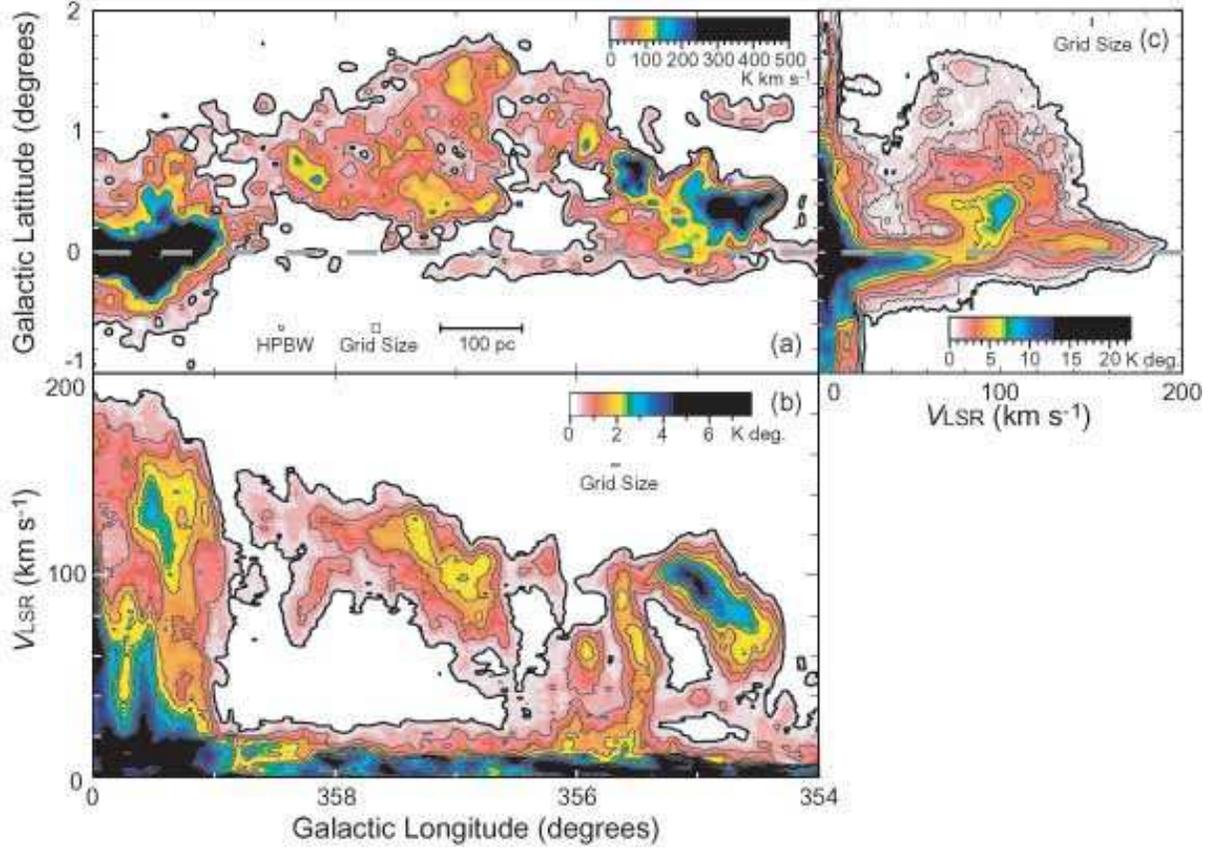}
  \end{center}
  \caption{Integrated intensity maps of \atom{C}{}{12}\atom{O}{}{} ($J=1$--$0$) emission.
           (a) Velocity-integrated intensity map. 
           The integrated velocity range is from 30 km s$^{-1}$ to 200 km s$^{-1}$.
           The lowest contour level is 10 K km s$^{-1}$ and denoted by the thick line.
           The contour intervals are 60 K km s$^{-1}$ 
           and 30 K km s$^{-1}$ level is plotted in addition.
           (b) Galactic longitude-velocity diagram.
           The integrated galactic latitude range is from $-0\fdg50$ to $2\fdg0$.
           The lowest contour level (thick line) and the contour intervals are 
           0.25 K deg and 0.50 K deg, respectively.
           (c) Velocity-galactic latitude diagram.
           The integrated galactic longitude range is from $354\fdg0$ to $0\fdg0$. 
           The lowest contour level is 0.58 K deg and denoted by the thick line.
           The contour intervals are 1.74 K deg 
           and 1.16 and 1.74 K deg levels are plotted in addition.
           The gray broken line in (a) and (c) indicate $b = 0^{\circ}$. 
           (Color Online)}
  \label{12COlbv}
\end{figure}

 Figure \ref{ent}b shows the large-scale distribution of the \atom{C}{}{12}\atom{O}{}{} ($J=1$--$0$) emission 
 in an area of $350^{\circ} \leq l \leq 10^{\circ}$ and $-5^{\circ} \leq b < 5^{\circ} $ in 20 km s$^{-1}$ $\leq V \leq 300$ km s$^{-1}$. 
 Figure \ref{12COlbv} focuses on an area of $354^{\circ} \leq l \leq 0^{\circ}$ and $-1^{\circ} \leq b \leq 2^{\circ}$. 
 We see that an elevated dome-like feature of $\sim4$ degree extent in $l$ 
 and $\sim2$ degree height in $b$ with relatively weak intensity 
 is dominant in the region (Figure \ref{12COlbv}a) 
 and this feature has a velocity gradient as seen in a longitude velocity diagram (Figure \ref{12COlbv}b). 
 The dome-like elevated feature is also recognized in latitude-velocity diagram (Figure \ref{12COlbv}c). 
 We here identify the dome-like feature as loop 3.
 
 \par
 
 The strong emission at $359^{\circ} \leq l \leq 0^{\circ}$ and $b = 0^{\circ}$ is part of the CMZ. 
 The feature at 0 km s$^{-1}$ $\leq V \leq 60$ km s$^{-1}$ is part of CMZ, 
 and the other at 110 km s$^{-1}$ $\leq V \leq 200$ km s$^{-1}$ is part of the expanding molecular ring (EMR) 
 (see \cite{mor1996}). 
 Clump 1 \citep{ban1986} is located at $354\fdg3 \leq l \leq 355\fdg3$ and $0^{\circ} \leq b \leq 1^{\circ}$. 
 In addition, we note three broad features of 50--100 km s$^{-1}$ 
 linewidths localized towards $(355\fdg4 \leq l \leq 355\fdg7,\ 0\fdg4 \leq b \leq 0\fdg9)$, $(355\fdg8 \leq l \leq 356\fdg0,\ 0\fdg7 \leq b \leq 1\fdg2)$ 
 and $(358\fdg0 \leq l \leq 358\fdg2,\ 0\fdg5 \leq b \leq 0\fdg9)$. 
 We identified such broad linewidth features as clumps using the following criteria: 
 (1) find a peak position in longitude-velocity diagram (Figure \ref{12COlbv}b), 
 (2) draw a contour at a level of two thirds of the peak integrated intensity, 
 (3) draw a contour at a 3 $\sigma$ noise level (10.4 K km s$^{-1}$) 
 in the longitude latitude map and identify contiguous points as a clump, 
 (4) exclude the position identified as a clump in procedure (3) and go back to procedure (1) 
 if further peak position exists. 
 By using these criteria, we have identified 5 clumps. 
 4 clumps are localized within 0\fdg4 towards longitude direction as listed 
 in Tables \ref{tab:obs_clumps} and \ref{tab:phys_clumps}.
 
 \par
 
 Broad linewidths such as 50 km s$^{-1}$ are much larger than that of disk molecular clouds. 
 They are also significant larger than those of the near and far sides of the 3 kpc arm of $\sim10$ km s$^{-1}$ 
 that are located at 3.5 kpc away from the center (\cite{dam2008}; see also \cite{ban1980}; 1986). 
 The large velocity spans of loop 3 are similar to the molecular gas in CMZ 
 rather than the 3 kpc arm molecular gas and we suggest that loop 3 is located near the Galactic center 
 and probably within $\sim1$ kpc of Sgr A$^{*}$. 
 Therefore, we assume that these molecular features are located in the Galactic center 
 and adopt a distance of 8.5 kpc, hereafter. 
 The projected length of loop 3 is $\sim600$ pc and the height is $\sim300$ pc from Galactic plane.
 
 \par
 
\begin{figure}
  \begin{center}
    \FigureFile(160mm,160mm){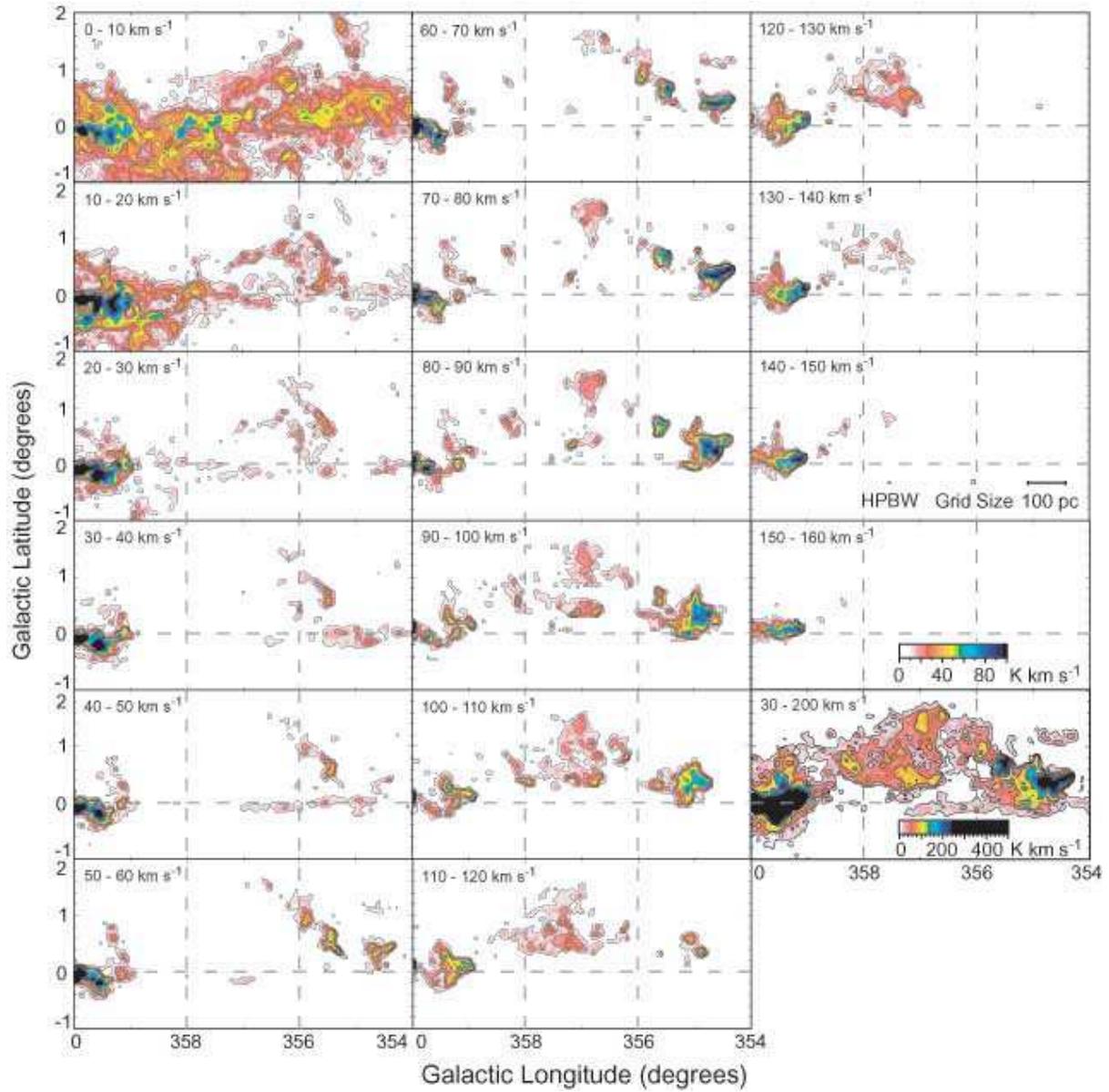}
  \end{center}
  \caption{Velocity channel maps of \atom{C}{}{12}\atom{O}{}{} ($J=1$--$0$) emission.
           The lowest contour level is 8.7 K km s$^{-1}$.
           The contour intervals are 10 K km s$^{-1}$ until 38.7 K km s$^{-1}$
           and then 20 K km s$^{-1}$. 
           The last panel is same as Figure \ref{12COlbv}a
           (Color Online)}
  \label{12COchannel}
\end{figure}

 More details of the \atom{C}{}{12}\atom{O}{}{} distribution are shown in Figure \ref{12COchannel} 
 as a series of velocity integrated distributions every 10 km s$^{-1}$. 
 We see that \atom{C}{}{12}\atom{O}{}{} emission in loop 3 shifts from the smaller longitude side 
 to the larger longitude side as the velocity becomes larger. 
 In addition, Figure \ref{12COchannel} shows two mini-loop features 
 at $356\fdg5 \leq l \leq 357\fdg5$ and $0\fdg2 \leq b \leq 1\fdg6$ in the 100--110 km s$^{-1}$ panel 
 and at $357\fdg0 \leq l \leq 358\fdg2$ and $0\fdg5 \leq b \leq 1\fdg3$ in the 120--130 km s$^{-1}$ panel.
 
 \par
 
\begin{figure}
  \begin{center}
    \FigureFile(160mm,95mm){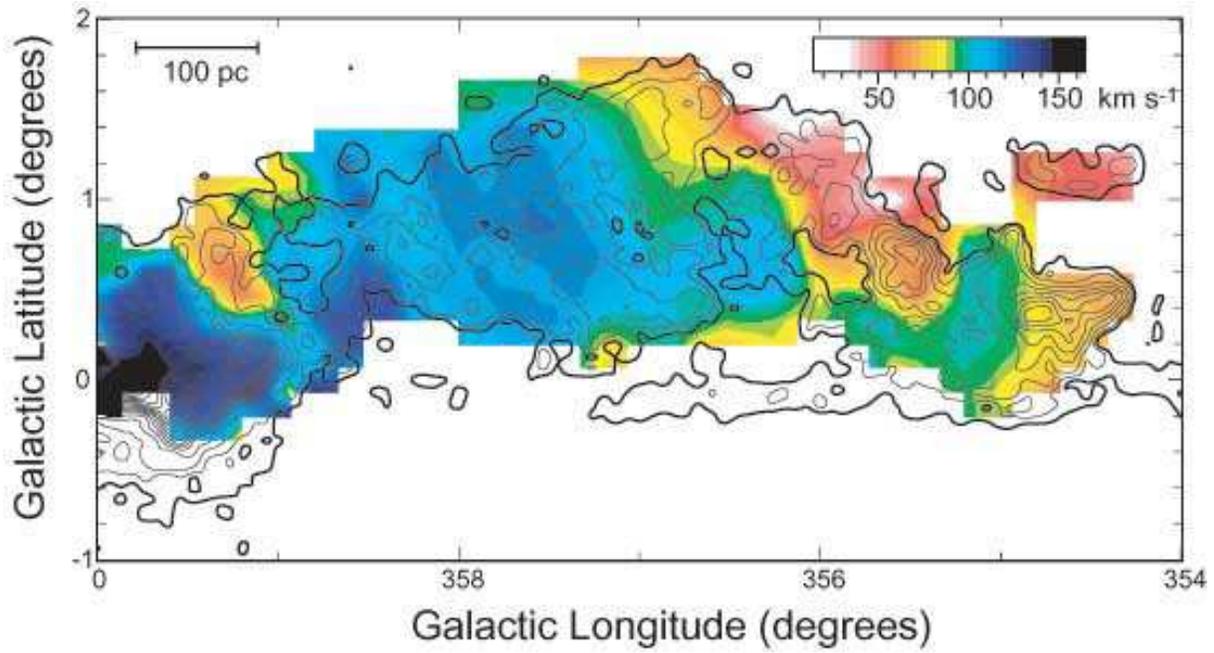}
  \end{center}
  \caption{Velocity centroid map of \atom{C}{}{12}\atom{O}{}{} ($J=1$--$0$) emission.
           Data are smoothed to 5\farcm2 beam size and $8\arcmin$ grid size.
           Contours are from Figure \ref{12COlbv}a. 
           (Color Online)}
  \label{12COvelo}
\end{figure}

 Figure \ref{12COvelo} shows a distribution of the velocity centroid of the \atom{C}{}{12}\atom{O}{}{} emission, 
 where the velocity centroid is the averaged velocity with intensity-weighting at each position. 
 This clearly indicates the velocity gradient over the dome-like feature at 30--130 km s$^{-1}$ 
 over $\sim3$ degrees. 
 Figure \ref{13COlbv} shows \atom{C}{}{13}\atom{O}{}{} ($J=1$--$0$) distribution. 
 The \atom{C}{}{13}\atom{O}{}{} is generally weak by a factor of 5--6 or more than \atom{C}{}{12}\atom{O}{}{}. 
 The \atom{C}{}{13}\atom{O}{}{} data were used to estimate the molecular mass in section 4.3.
  
\subsection{Comparison with H\emissiontype{I}}
 We next compare the CO with H\emissiontype{I} in Figure \ref{HIlbv}. 
 The same CO distributions in Figure \ref{12COlbv} are overlaid with the H\emissiontype{I} distribution 
 at a $16\arcmin$ effective resolution taken with the Parkes 64 m telescope \citep{mcc2005}. 
 The H\emissiontype{I} distribution in intensity and velocity are 
 basically consistent with CO while the H\emissiontype{I} is more continuously distributed 
 than the CO and is apparently enveloping the CO. 

\subsection{Mass Estimates}

 The total molecular mass of loop 3 is estimated using the X-factor, 
 which is the empirical conversion factor from \atom{C}{}{12}\atom{O}{}{} integrated intensity to molecular hydrogen column density:
 \begin{eqnarray}
  X = N(\mathrm{H}_{2}) / W 
 \end{eqnarray}
 where $N(\mathrm{H}_{2})$ is the molecular hydrogen column density in cm$^{-2}$ units 
 and $W$ is the \atom{C}{}{12}\atom{O}{}{} integrated intensity in K km s$^{-1}$ units. 
 As we see in Figure \ref{12COlbv}b and c, \atom{C}{}{12}\atom{O}{}{} emission is distributed uniformly 
 in the region of $V \leq 30$ km s$^{-1}$. 
 In this low velocity range, the foreground emissions are significant 
 and the integrated range is set to 30 km s$^{-1}$ $\leq V \leq 200$ km s$^{-1}$ to extract loop 3 
 and related features in the Galactic center in the present study. 
 The mass of the molecular gas is calculated using following equation:
 \begin{eqnarray}
	M(\mathrm{\atom{C}{}{12}\atom{O}{}{}}) = u m_{\mathrm{H}} \sum^{}_{} \bigl[ D^{2} \Omega N(\mathrm{H}_{2}) \bigr]
 \end{eqnarray}
 where $u$ is the mean molecular weight which is assumed to be 2.8, 
 $m_{\mathrm{H}}$ is the atomic hydrogen mass, 
 $D$ is the distance of loop 3, 8.5 kpc, 
 and $\Omega$ is the solid angle subtended by a unit grid spacing $4\arcmin \times 4\arcmin$. 
 Two values of the X-factor, $0.24 \times 10^{20}$ cm$^{-2}$ (K km s$^{-1}$)$^{-1}$ \citep{oka1998a} 
 and $0.6 \times 10^{20}$ cm$^{-2}$ (K km s$^{-1}$)$^{-1}$ \citep{oni2004}, were adopted. 
 The summations were performed over the observed points 
 within the 3 $\sigma$ contour level of the integrated intensity 
 by excluding the CMZ and the Clump 1 regions, 
 and the total molecular mass of loop 3 is calculated to be 
 $1.9 \times 10^{6} \Mo$ and $4.8 \times 10^{6} \Mo$, respectively, for the two X-factors.
 
 \par
 
 Although \atom{C}{}{13}\atom{O}{}{} observations are confined to $b < 1^{\circ}$, 
 \atom{C}{}{13}\atom{O}{}{} column densities are calculated by assuming the local thermodynamic equilibrium (LTE) 
 to estimate the lower limit of the molecular mass. 
 The optical depth of \atom{C}{}{13}\atom{O}{}{}, $\tau(\mathrm{\atom{C}{}{13}\atom{O}{}{}})$, was calculated using following equation:
 \begin{eqnarray}
       \tau(\mathrm{\atom{C}{}{13}\atom{O}{}{}}) = - \ln \left[ 1 - \frac{T^{*}_{\mathrm{R}}(\mathrm{\atom{C}{}{13}\atom{O}{}{}})}{5.29 \times (J(T_{\mathrm{ex}})-0.164)} \right]
 \end{eqnarray}
 where $T^{*}_{\mathrm{R}}(\mathrm{\atom{C}{}{13}\atom{O}{}{}})$ and $T_{\mathrm{ex}}$ 
 are the radiation temperature and the excitation temperature of \atom{C}{}{13}\atom{O}{}{}, respectively. 
 $J(T)$ is defined as $J(T) = 1 / \left[ \exp(5.29/T)-1 \right]$. $N(\mathrm{\atom{C}{}{13}\atom{O}{}{}})$ was estimated from
 \begin{eqnarray}
       N(\mathrm{\atom{C}{}{13}\atom{O}{}{}}) = 2.42 \times 10^{14} \frac{\tau(\mathrm{\atom{C}{}{13}\atom{O}{}{}}) \Delta V T_{\mathrm{ex}}}{1-\exp(-5.29/T_{\mathrm{ex}}))}
 \end{eqnarray}
 where $\Delta V$ is the \atom{C}{}{13}\atom{O}{}{} linewidth. $N(\mathrm{\atom{C}{}{13}\atom{O}{}{}})$ is calculated 
 for each observed position which matches the observed position in \atom{C}{}{12}\atom{O}{}{}. 
 In this study, \atom{C}{}{12}\atom{O}{}{} peak temperatures are adapted for $T_{\mathrm{ex}}$, 
 because \atom{C}{}{12}\atom{O}{}{} is usually optically thick. 
 $N(\mathrm{H}_{2})$ is estimated by assuming that the [H$_{2}$]/[\atom{C}{}{13}\atom{O}{}{}] ratio is $1 \times 10^{6}$ \citep{lis1989}
 which is the value of Sgr B2.
 The summations were performed in the same way as \atom{C}{}{12}\atom{O}{}{}, 
 and total mass was derived to be $1.7 \times 10^{6} \Mo$. 
 If we assume that the [H$_{2}$]/[\atom{C}{}{13}\atom{O}{}{}] ratio is $5 \times 10^{5}$ \citep{dic1978}
 which is the average value of local clouds, total mass is calculated to be $8.3 \times 10^{5} \Mo$.
 
 \par
 
 H\emissiontype{I} column densities were calculated assuming that the H\emissiontype{I} 21 cm line is optically thin 
 and $T_{\mathrm{s}} \gg h\nu/k$. The following equation was used.
 \begin{eqnarray}
	N(\mathrm{H}) = 1.82 \times 10^{18} \int T dv 
 \end{eqnarray}
 In \atom{C}{}{12}\atom{O}{}{} we set an integration range as $30$ km s$^{-1}$ $\leq V \leq$ $200$ km s$^{-1}$, 
 but the foreground component is significant in $\leq$ 50 km s$^{-1}$ for H\emissiontype{I} gas. 
 Therefore, we calculated the H\emissiontype{I} column density by setting an integration range of 
 $50$ km s$^{-1}$ $\leq V \leq$ $200$ km s$^{-1}$. 
 The summations were performed by the same as \atom{C}{}{12}\atom{O}{}{}, and total mass was derived to be $1.4 \times 10^{6} \Mo$.
 
 \par
 
 The cloud masses estimated in \atom{C}{}{12}\atom{O}{}{}, \atom{C}{}{13}\atom{O}{}{} and H\emissiontype{I} for loop 3 are summarized in Table \ref{tab:mass}. 
 The H\emissiontype{I} and the molecular masses are comparable. 
 By adopting each of the molecular and atomic masses is $\sim1.5 \times 10^{6} \Mo$, 
 the total gaseous mass in loop 3 is derived as $\sim3.0 \times 10^{6} \Mo$.

\newpage
 
\section{Discussion}

\subsection{Physical Properties of Loop 3}
 
\begin{figure}
  \begin{center}
    \FigureFile(80mm,140mm){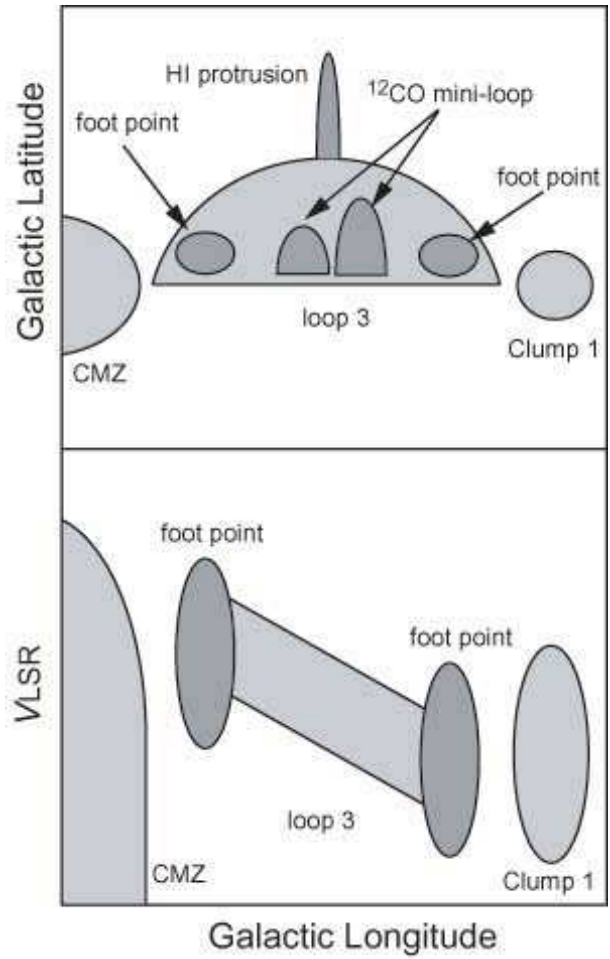}
  \end{center}
  \caption{Schematic views of the velocity integrated map and the Galactic longitude-velocity map.}
  \label{sum}
\end{figure}

 Figure \ref{12COlbv}b shows that the molecular gas in loop 3 has broad spans of 
 $\sim100$ km s$^{-1}$ in the both ends at $l \sim 355\fdg5$ and $\sim358\fdg3$ 
 and they are the broad features a and d in Tables \ref{tab:obs_clumps} and \ref{tab:phys_clumps}. 
 We identify the two features as the foot points of loop 3 
 by considering the compact and broad properties similar to those of loops 1 and 2. 
 The nearly monotonic velocity gradient between the both ends, $\sim80$ km s$^{-1}$ over $\sim2$ degrees 
 in $l$, is also consistent with the kinetic properties of loops 1 and 2 (Figure \ref{12COvelo}). 
 These properties meet the observational properties of a magnetic loop described  in section 3.2 
 but do not meet the properties of a feature created by a supershell 
 due to multiple supernova explosions; 
 supernova explosions create a curved feature but not a feature with a uniform velocity gradient 
 in the Galactic longitude-velocity diagram shown in Figure \ref{12COlbv}b \citep{fuk2006}. 
 Figure \ref{sum} gives a schematic view of loop 3. 
 
 \par
 
 The distance of loop 3 from the Galactic center is not certain 
 because of the unknown kinematics in the nuclear gas disk. 
 The position of loop 3 in the velocity galactic longitude diagram 
 is apparently forbidden in pure rotation and some radial motion 
 either expansion or contraction in the order of 100 km s$^{-1}$ is definitely required. 
 If we assume that loop 3 is located at a radius from the Galactic center, 
 the projected outer longitude, $-5^{\circ}$, poses a lower limit 
 in the radius to be $\sim750$ pc. 
 Since the magnetic instability requires a large field strength under a strong stellar gravity, 
 loop 3 must be within 1.2 kpc of the center where the stellar gravity force 
 is stronger than that in the solar vicinity by a factor of 10. 
 We thus tentatively assume that loop 3 is located along a ring of $\sim1$ kpc radius from the center. 
 More discussion will be found in section 5.4.
 
 \par
 
 The kinetic energy involved in loop 3 is estimated to be $\sim1.7 \times 10^{52}$ ergs 
 for a total mass of $\sim3.0 \times 10^{6} \Mo$ and a velocity dispersion of $\sim30$ km s$^{-1}$, 
 the average velocity dispersion of all the observed positions in loop 3. 
 This kinetic energy is too large to be explained by a single supernova explosion and the spatial extent, 
 $\sim600$ pc, is too large to be explained by a star cluster to form a supershell. 
 This is consistent with an interpretation that loop 3 is due to magnetic floatation by the Parker instability, 
 the same mechanism for loops 1 and 2.
 
 \par
 
 If we assume the energy equi-partition between molecular gas kinetic energy 
 and magnetic field energy following \citet{fuk2006}, 
 the equation of energy conservation is written as;
 \begin{eqnarray}
	\frac{1}{2} \mu m_{\mathrm{H}} n \Delta V^{2} = \frac{B^{2}}{8 \pi}
 \end{eqnarray}
 where $\mu$ is mean molecular weight including molecular hydrogen and 20\% 
 of helium atom in number, $n$ is density of neutral particles, 
 $\Delta V$ is the turbulent velocity of molecular gas, 
 and $B$ is the magnetic field. Magnetic field is estimated as $\sim150$ $\mu$G 
 by taking $n \sim 100$ cm$^{-3}$ and $\Delta V \sim 30$ km s$^{-1}$, similar values with loops 1 and 2. 
 The Alfv$\mathrm{\acute{e}}$n speed, $V_{\mathrm{A}}$, 
 is estimated as $V_{\mathrm{A}} = B / \sqrt{4 \pi \rho} = 24$ km s$^{-1}$. 
 Magnetic loops typically rise up at a velocity of $\sim V_{\mathrm{A}}$(Matsumoto et al. 1988), 
 and the time scale of this flotation, $t_{\mathrm{loop3}}$ is derived as 
 $t_{\mathrm{loop3}} = h / V_{\mathrm{A}} \simeq 10^7$ yr, where $h$ is the height of loop 3.
 
\subsection{Comparison with Loops 1 and 2}
 
 Physical properties of loops 1, 2 and 3 are presented in Table \ref{tab:loops}. 
 Loop 3 is similar to loops 1 and 2 in the height and the length 
 but is different from loops 1 and 2 in the morphology; 
 loops 1 and 2 show clear loop-like features whose inside show no molecular emission, 
 while loop 3 shows a dome-like feature whose inside is filled with the molecular emission.
 
 \par
 
 Another difference is the relative weakness in intensity of the foot points in loop 3 
 as compared to loops 1 and 2. 
 Foot points of a loop are formed by the falling down motion of the gas onto the galactic plane 
 \citep{mat1988}. 
 In order to measure the degree of mass-wise difference of the foot points 
 we have estimated the line intensity ratios 
 between the foot point regions and the entire loop as listed in col.~(8) of Table \ref{tab:loops}. 
 The ratios of loops 1 and 2 are calculated together 
 because loops 1 and 2 are overlapped at the foot point, 
 making it difficult to separate each contribution. 
 Table \ref{tab:loops} shows that the ratio of loop 3 is about a half of that of loops 1 and 2.
 
 \par
 
 When we consider the intensity ratio of the foot point, the curvature of a magnetic loop is also important. 
 For the gas in a magnetic loop having a smaller curvature or flatter top, 
 the effective gravity for the gas is less. 
 Therefore, the gas in such a magnetic loop tends to remain in the loop 
 and one expects the loop shows a lower intensity ratio. 
 It is uncertain to compare curvatures of loops 1, 2 and 3 quantitatively due to the projection effect. 
 The projected length of loop 3 is twice as large as that of loop 1 and 4 times larger than that of loop 2. 
 This might suggest that loop 3 has a smaller curvature and offer another explanation 
 that the foot points of loop 3 are less developed than those of loops 1 and 2.
 
 \par
 
 Finally in this subsection, we suggest that Clump 1 is another candidate for a magnetic loop 
 as suggested by its velocity gradient. 
 It may be a young one in the earliest phase of floatation. 
 Searches for similar small loops must be of considerable interest 
 in order to shed light on the initiation of flotation.
 
\subsection{H\emissiontype{I} Protrusion}
 
 Figure \ref{HIlbv}a shows that the protrusion of the H\emissiontype{I} gas 
 is located in $357^{\circ} \leq l \leq 358^{\circ}$ and $2^{\circ} \leq b \leq 4^{\circ}$. 
 This feature has velocity of $\sim100$ km s$^{-1}$ and is not a foreground component. 
 The positions of the H\emissiontype{I} protrusion meets the center of loop 3 at $(l,\ b) \sim (357\fdg5,\ 2^{\circ})$ 
 and is within the velocity range of loop 3. 
 Therefore, we suggest the H\emissiontype{I} protrusion is possibly associated with loop 3. 
 The H\emissiontype{I} protrusion has no \atom{C}{}{}\atom{O}{}{} counterpart, implying its low density.
 
 \par
 
\begin{figure}
  \begin{center}
    \FigureFile(160mm,70mm){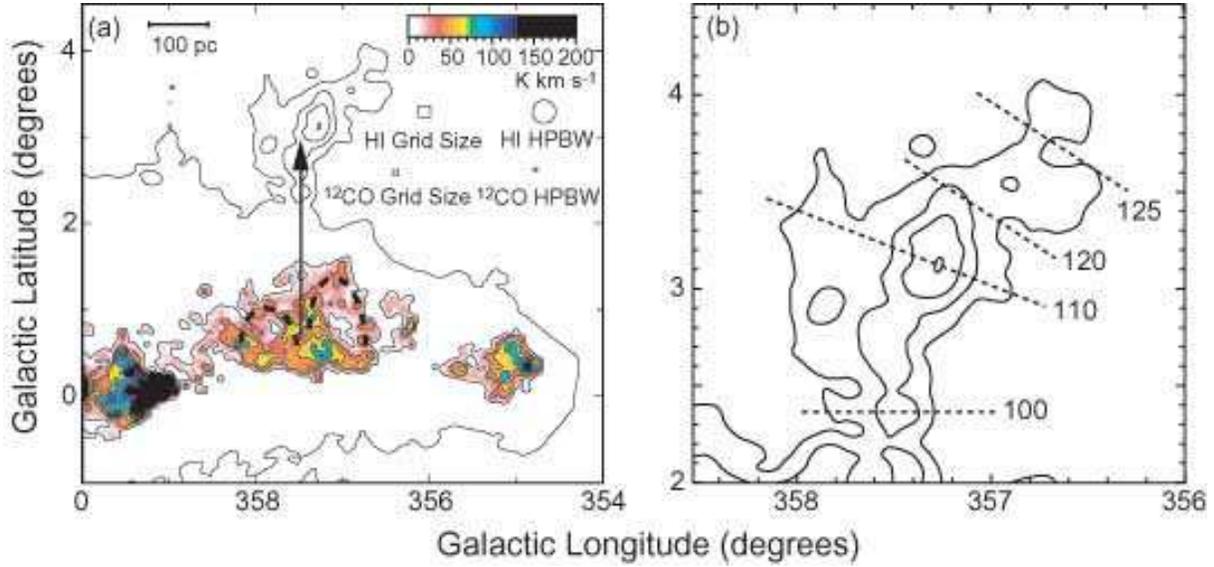}
  \end{center}
  \caption{(a) Velocity-integrated intensity map of H\emissiontype{I} protrusion and \atom{C}{}{12}\atom{O}{}{} mini-loops. 
           H\emissiontype{I}: The integrated velocity range is from 80 km s$^{-1}$ to 170 km s$^{-1}$.
           The lowest contour level and the contour intervals are 
           21 K km s$^{-1}$ and 20 K km s$^{-1}$, respectively. 
           Only the lowest contour and the protrusion componet are plotted.
           \atom{C}{}{12}\atom{O}{}{} ($J=1$--$0$): 
           The integrated velocity range is from 100 km s$^{-1}$ to 140 km s$^{-1}$
           The lowest contour level and the contour intervals are 
           10.4 K km s$^{-1}$ and 20.8 K km s$^{-1}$, respectively. 
           The dashed lines indicate two \atom{C}{}{12}\atom{O}{}{} mini-loops and 
           the arrow indicate the H\emissiontype{I} protrusion trajectory.
           (b) Close-up view of H\emissiontype{I} protrusion. The contour levels are same as (a)
           Numbers indicate the LSR velocities in km s$^{-1}$. 
           (Color Online)}
  \label{jetminiloop}
\end{figure}

 Figure \ref{jetminiloop}a shows the present H\emissiontype{I} protrusion resolved with a $16\arcmin$ beam. 
 The protrusion is confined to $\sim 1^{\circ}$ towards Galactic longitude of $\sim 357^{\circ}$ 
 and elongated by $\sim 2^{\circ}$ in Galactic latitude nearly perpendicular to the Galactic plane. 
 We also note that the H\emissiontype{I} protrusion shows a velocity gradient towards the northern top in Figure \ref{jetminiloop}b.
 
 \par
 
 It is interesting to note that the two \atom{C}{}{12}\atom{O}{}{} mini-loops are seen in a velocity range 
 around 100--120 km s$^{-1}$ in Figure \ref{12COchannel}, which may be parts of loop 3. 
 The multiple broad features of $\sim50$ km s$^{-1}$, c and d listed in Tables \ref{tab:obs_clumps} and \ref{tab:phys_clumps}, 
 may represent the foot points of these loops. 
 Multiple loops are in fact formed simultaneously on the surface of the Sun (e.g., \cite{you2007}). 
 If the H\emissiontype{I} protrusion trajectory is extrapolated to the lower side along the Galactic longitude, 
 the trajectory coincides with the "valley" between the two mini-loops (Figure \ref{jetminiloop}a). 
 We may speculate that the origin of the protrusion is magnetic reconnection between the mini-loops or spur (e.g., \cite{mat1988}). 
 Magnetic reconnections likely occur at a valley between the magnetic loops 
 because the magnetic fields have opposite directions and in contact with each other. 
 The velocity gradient in Figure \ref{jetminiloop}b is consistent with the deceleration 
 in the ejection scenario from the Galactic plane either 
 by the magnetic reconnection or spur since the acceleration does not work 
 at higher latitudes where the magnetic field becomes weaker.
 
 \par
 
\begin{figure}
  \begin{center}
    \FigureFile(80mm,60mm){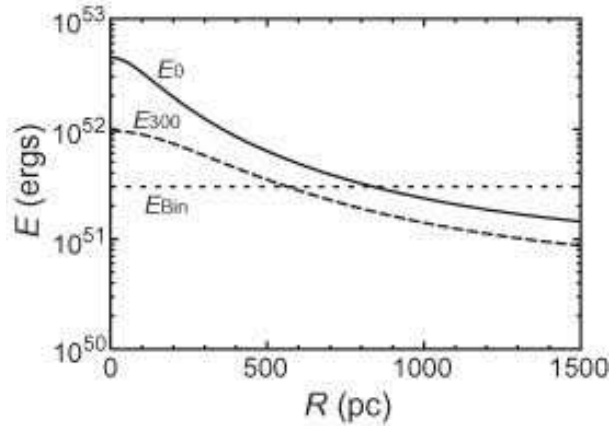}
  \end{center}
  \caption{Single logarithmic plot of the energy versus the distance from galactic rotation axis.
           The solid line, the broken line and the dashed line are 
           the gravitational potential energy from $z = 0$ pc to $z = 500$ pc,
           the gravitational potential energy from $z = 300$ pc to $z = 500$ pc and
           the magnetic energy, respectively.}
  \label{E-R}
\end{figure}

 We shall here compare the energy of the magnetic fields and the H\emissiontype{I} protrusion. 
 \atom{C}{}{12}\atom{O}{}{} mini-loops have a width of 40 pc and a height of 100 pc. 
 Therefore, we consider the volume in which magnetic reconnection occurs as 
 40 pc $\times$ 40 pc $\times$ 100 pc. 
 If we assume that the magnetic field of 150 $\mu$G flow into the space 
 at a velocity of $0.1 \times V_{\mathrm{A}}$ in a time scale of $t_{\mathrm{loop3}}$, 
 then the energy of magnetic field is estimated at $E_{B_\mathrm{in}} \simeq 3 \times 10^{51}$ ergs. 
 On the other hand, energy of H\emissiontype{I} protrusion under the influence of the gravitational potential 
 is calculated using Miyamoto-Nagai potential modified by \citet{sof1996}. 
 The equation is expressed by $(R, z)$ coordinates where $R$ and $z$ is 
 the distance from galactic rotation axis and galactic plane, respectively, 
 and consists of 4 mass components as follows:
 \begin{eqnarray}
     \Phi(R,\ z) = \sum^{4}_{\mathrm{i}=1} 
                   \left[ 
                   \frac{ G M_{\mathrm{i}} }
                        { (R^{2}+(a_{\mathrm{i}} + (z^{2}+b_{\mathrm{i}}^{2})^{\frac{1}{2}})^2)^{\frac{1}{2}} } 
                   \right]
 \end{eqnarray}
 where $M_{\mathrm{i}}$, $a_{\mathrm{i}}$, and $b_{\mathrm{i}}$ are 
 the mass, scale radius, and scale thickness of ith mass component. 
 $M_{\mathrm{i}}$, $a_{\mathrm{i}}$, and $b_{\mathrm{i}}$ fitted by \citet{sof1996} are summarized in Table \ref{tab:pot}. 
 H\emissiontype{I} protrusion mass, $M_{\mathrm{prot}}$, is estimated at $1.5 \times 10^{4} \Mo$. 
 The integrated intensity peak is located at $b \sim 3^{\circ}$, 
 and therefore we consider the energy that a H\emissiontype{I} cloud of $1.5 \times 10^{4} \Mo$ 
 is lifted up to 500 pc from Galactic plane under the influence of gravity. 
 We consider the two cases, the H\emissiontype{I} gas is lifted up from $z = 0$ pc and $300$ pc, 
 i.e., $ M_{\mathrm{prot}} (\Phi(R,\ 0) - \Phi(R,\ 500)) \equiv E_{0}(R)$ 
 and $ M_{\mathrm{prot}} (\Phi(R,\ 300) - \Phi(R,\ 500)) \equiv E_{300}(R)$. 
 Figure \ref{E-R} shows plots of $E_{0}(R)$, $E_{300}(R)$ and $E_{B_\mathrm{in}}$. 
 In the case of $E_{0}(R)$, $E_{B_\mathrm{in}}$ exceeds the gravitational potential energy 
 in the region of $R \gtrsim 800$ pc. 
 In the case of $E_{300}(R)$, $E_{B_\mathrm{in}}$ exceeds in the region of $R \gtrsim 600$ pc. 
 In such cases, if part of the magnetic energy is converted into the H\emissiontype{I} protrusion kinetic energy 
 through magnetic reconnections, 
 the reconnections are able to explain the energy of the H\emissiontype{I} protrusion. 
 The following discussion in section 5.4 suggests that loop 3 is located 
 at a radius from 750 pc to 1.2 kpc from the center.
 
\subsection{Kinematics and Location of Loop 3}
 
 The kinematics with non-circular motions in the nuclear gas disk is explained 
 in terms of the stellar bar driven potential \citep{bin1991}. 
 This is a viable way to create general non-circular motions in the Galactic center 
 as demonstrated by numerical simulations by these authors. 
 The magnetic flotation is also able to create radial motions 
 as a result of magneto-rotational instability followed by the Parker instability \citep{mac2009}. 
 In fact, the magnetic instability should work also in the bar-like gravitational potential 
 and the two models are not exclusive. 
 It is desirable in future to test how the magnetic instability works on the gas dynamics in a bar-like potential.
 
 \par
 
 The global numerical simulations of magnetic instabilities by \citet{mac2009} 
 indicates that the $m = 1$ mode becomes dominant in the nuclear gas disk. 
 The three loops 1, 2 and 3 are all located in the negative galactic longitude side 
 of the center whereas the CMZ, the densest molecular aggregation, 
 is located in the opposite side of the center. 
 We note that this global molecular distribution is consistent with the results of \citet{mac2009}.
 
 \par
 
 All the molecular gas of loop 3 has positive velocities. 
 If we assume that loop 3 is on part of a circle centered at Sgr A$^{*}$, 
 we are able to constrain the radius to be 750 pc to 1.2 kpc 
 by considering that loop 3 have to be within $\sim1$ kpc 
 where the stellar gravity force is high enough for the strong magnetic field of $\sim0.1$ mG 
 required for the Parker instability and by its projected position. 
 Observations also support to locate loop 3 within $\sim1$ kpc of the center, 
 since there are no such broad features outside 8 degrees in Galactic longitude in the NANTEN GPS dataset. 
 We then estimate the velocity of loop 3
 on the assumption that loop 3 is part of a circle centered at the Galactic center.
 The radius of the circle is assumed to be 1 kpc. 
 We also assume that loop 3 is uniformly expanding but is not contracting 
 by considering the expanding motion of the molecular ring, EMR. 
 The velocity with respect to the LSR is then expressed as follows for $l$;
 \begin{eqnarray}
      V &=&   V_{\mathrm{rot}} \frac{R_{0}}{R} \sin{l}
        +   V_{\mathrm{exp}}   \left( 1 - \frac{R_{0}^{2}}{R^{2}} \sin^{2}{l}  \right) ^{\frac{1}{2}}
        \pm V_{\mathrm{fall}}  \cos{a} \frac{R_{0}}{R} \sin{l} 
        -   V_{\mathrm{sun}} \sin{l}
 \end{eqnarray}
 where $R_{0}$ is the distance of the sun from the Galactic center of 8.5 kpc, 
 $R$ is the distance of loop 3 from the Galactic center, 
 $V_{\mathrm{rot}}$ is the rotation velocity, 
 $V_{\mathrm{exp}}$ is the expansion velocity, 
 $V_{\mathrm{fall}}$ is the falling velocity along the loop, 
 $V_{\mathrm{sun}}$  is the rotation velocity of the LSR about the center of 220 km s$^{-1}$, 
 and $a$ is the angle between the loop and the galactic plane at the foot point.
 This relation is solved for the two conditions at the two foot points that 
 $(l,\ V) = (359^{\circ},\ 130$ km s$^{-1})$ and $(355^{\circ},\ 20$ km s$^{-1})$ 
 assuming that $V_{\mathrm{fall}} \simeq V_{\mathrm{A}}$ at foot points \citep{mat1988}. 
 With $V_{\mathrm{A}}$ of 24 km s$^{-1}$, a rotational velocity and an expansion velocity 
 are derived as $\sim80$ km s$^{-1}$ and $\sim130$ km s$^{-1}$, respectively.
 
\newpage

\section{Conclusions}
 We have discovered a molecular dome-like feature towards $355^{\circ} \leq l \leq 359^{\circ}$ in $0^{\circ} \leq b \leq 2^{\circ}$. 
 The large velocity dispersions, 50--100 km s$^{-1}$, of the feature indicate 
 that it is located in the Galactic center, most probably within $\sim1$ kpc of Sgr A$^{*}$. 
 The distribution has a projected length of $\sim600$ pc and height of $\sim300$ pc from the Galactic disk 
 and shows a large-scale velocity gradient of $\sim130$ km s$^{-1}$ per $\sim600$ pc. 
 The feature is also associated with the H\emissiontype{I} gas having a more continuous spatial and velocity distribution than that of \atom{C}{}{12}\atom{O}{}{}. 
 We interpret the feature as one of the magnetically floated loops, 
 similar to the first two magnetically floated loops 1 and 2 \citep{fuk2006}, and name it "loop 3". 
 Loop 3 shares common observed properties in its height and length with loops 1 and 2. 
 We have also identified two foot points having broad linewidths at the both ends of loop 3, 
 and they are a typical observational signature of a magnetically floated loop. 
 It seems different from loops 1 and 2 in shape since the inner part of loop 3 is filled with molecular emission. 
 H\emissiontype{I}, \atom{C}{}{12}\atom{O}{}{} and \atom{C}{}{13}\atom{O}{}{} datasets were used to estimate the total mass and kinetic energy of loop 3 
 to be $\sim3.0 \times 10^{6} \Mo$ and $\sim1.7 \times 10^{52}$ ergs. 
 The huge size, velocity dispersions and energy are consistent with the magnetic flotation 
 as in case of loops 1 and 2 but are difficult to be explained by multiple stellar explosions.  
 We argue that loop 3 is in an earlier evolutionary phase than loops 1 and 2, 
 where the loop still has rising components from the disk and the shape of the loop top 
 is not well established like in loops 1 and 2. 
 This discovery indicates that the negative longitude side of Sgr A$^{*}$ 
 is dominated by three well developed magnetically floated loops.
 
 \bigskip
 
 We greatly appreciate the hospitality of all staff members
 of the Las Campanas Observatory of the Carnegie Institution
 of Washington. The NANTEN telescope was operated based
 on a mutual agreement between Nagoya University and the
 Carnegie Institution of Washington. We also acknowledge
 that the operation of NANTEN can be realized by contributions
 from many Japanese public donators and companies.
 This work is financially supported in part by a Grant-in-Aid
 for Scientific Research (KAKENHI) from the Ministry
 of Education, Culture, Sports, Science and Technology of
 Japan (Nos.~15071203 and 18026004) and from JSPS (Nos.~14102003, 20244014, and 18684003). 
 This work is also financially
 supported in part by core-to-core program of a Grant-in-Aid 
 for Scientific Research from the Ministry of Education,
 Culture, Sports, Science and Technology of Japan (No.~17004).

\newpage


\newpage

\begin{table}
  \begin{center}
 \caption{Observed properties of \atom{C}{}{12}\atom{O}{}{} broad linewidth features}\label{tab:obs_clumps}
    \begin{tabular}{cllllllrrrrr}

\hline \hline 
  & & & & & \multicolumn{4}{c}{Temperature peak position} & & & \\
  \cline{6-9}

\multicolumn{1}{c}{Name} & \multicolumn{1}{c}{$l_{\mathrm{min}}$} & \multicolumn{1}{c}{$l_{\mathrm{max}}$} &
\multicolumn{1}{c}{$b_{\mathrm{min}}$} & \multicolumn{1}{c}{$b_{\mathrm{max}}$} &
\multicolumn{1}{c}{$l$} & \multicolumn{1}{c}{$b$} & \multicolumn{1}{c}{$V_{\mathrm{LSR}}$} & 
\multicolumn{1}{c}{$T_{\mathrm{R}}^{*}$} & 
\multicolumn{1}{c}{$V_{\mathrm{LSR,mean}}$\footnotemark[$\dagger$]} & 
\multicolumn{1}{c}{$\Delta V$} & \multicolumn{1}{c}{coments} \\

\multicolumn{1}{c}{} & \multicolumn{1}{c}{($^{\circ}$)} & \multicolumn{1}{c}{($^{\circ}$)} & 
\multicolumn{1}{c}{($^{\circ}$)} & \multicolumn{1}{c}{($^{\circ}$)} & 
\multicolumn{1}{c}{($^{\circ}$)} & \multicolumn{1}{c}{($^{\circ}$)} & \multicolumn{1}{c}{(km\ s$^{-1}$)} & 
\multicolumn{1}{c}{(K)} & \multicolumn{1}{c}{(km\ s$^{-1}$)} & 
\multicolumn{1}{c}{(km\ s$^{-1}$)} & \multicolumn{1}{c}{} \\

\hline

a & 355.40 & 355.67 & 0.40 & 0.87 & 355.47 & 0.67  & 67 & 9.5  & 72 & 54 & MGC 355.5-0.7\footnotemark[$\ddagger$] \\
b & 355.87 & 355.93 & 0.73 & 1.33 & 355.93 & 0.87  & 62 & 6.4  & 62 & 13 & \\
c & 356.20 & 356.27 & 0.67 & 1.13 & 356.20 & 0.80  & 105 & 5.5  & 106 & 10 & \\
d & 357.93 & 358.27 & 0.33 & 0.93 & 357.93 & 0.40  & 104 & 5.1  & 97 & 27 & \\

\hline

\multicolumn{12}{@{}l@{}}{\hbox to 0pt{\parbox{180mm}{
  \par\noindent
  \footnotemark[$\dagger$] Mean LSR velocity weighed by the intensity.
  \par\noindent
  \footnotemark[$\ddagger$] Named in the present study.

}\hss}}

    \end{tabular}
  \end{center}
\end{table}

\newpage

\begin{table}
  \begin{center}
 \caption{Physical properties of \atom{C}{}{12}\atom{O}{}{} broad linewidth features}\label{tab:phys_clumps}
    \begin{tabular}{crr}

\hline \hline 

\multicolumn{1}{c}{Name} & \multicolumn{1}{c}{$R$\footnotemark[$*$]} 
& \multicolumn{1}{c}{$M$\footnotemark[$\dagger$]} \\

\multicolumn{1}{c}{} & \multicolumn{1}{c}{(pc)}
& \multicolumn{1}{c}{($10^{5}\Mo$)}\\

\hline

a & 32 & 1.4--3.4 \\
b & 23 & 0.33--0.83 \\
c & 20 & 0.14--0.36 \\
d & 28 & 0.31--0.78 \\

\hline

\multicolumn{3}{@{}l@{}}{\hbox to 0pt{\parbox{85mm}{
  \par\noindent
  \footnotemark[$*$]  Effective radii defined as $R=(A/\pi)^{0.5}$, where $A$ is the total cloud surface area. 
                      The distance is assumed to be 8.5 kpc.
                    
  \par\noindent
  \footnotemark[$\dagger$]  Calculated using X-factors of $0.24 \times 10^{20}$ cm$^{-2}$ (K km s$^{-1}$)$^{-1}$ 
                    (Oka et al. 1998a) for lower value and $0.6 \times 10^{20}$ (cm$^{-2}$ (K km s$^{-1}$)$^{-1}$)
                    (Onishi et al. 2004) for higher value.
                    
}\hss}}

    \end{tabular}
  \end{center}
\end{table}

\newpage

\begin{table}
  \begin{center}
 \caption{\atom{C}{}{12}\atom{O}{}{}, \atom{C}{}{13}\atom{O}{}{} and H\emissiontype{I} cloud mass}\label{tab:mass}
    \begin{tabular}{rlr}

\hline \hline 

\multicolumn{1}{c}{$X$} & \multicolumn{1}{c}{reference} & \multicolumn{1}{c}{$M(\mathrm{\atom{C}{}{12}\atom{O}{}{}})$} \\

\multicolumn{1}{c}{(cm$^{-2}$ (K km s$^{-1}$)$^{-1}$)} & \multicolumn{1}{c}{} & \multicolumn{1}{c}{($\Mo$)} \\

\hline

$0.24 \times 10^{20}$ & Oka et al. 1998a    & $1.9  \times 10^{6}$ \\
$0.6 \times 10^{20}$  & Onishi et al. 2004 & $4.8  \times 10^{6}$ \\

\hline
\hline

\multicolumn{1}{c}{[H$_{2}$]/[\atom{C}{}{13}\atom{O}{}{}]} & \multicolumn{1}{c}{reference} & \multicolumn{1}{c}{$M(\mathrm{\atom{C}{}{13}\atom{O}{}{}})$} \\

\multicolumn{1}{c}{} & \multicolumn{1}{c}{} & \multicolumn{1}{c}{($\Mo$)} \\

\hline

$5 \times 10^{5}$ & Dickman 1978           & $0.83  \times 10^{6}$ \\
$1 \times 10^{6}$ & Lis and Goldsmith 1989 & $1.7  \times 10^{6}$ \\

\hline
\hline

\multicolumn{1}{c}{Integrated range} & \multicolumn{1}{c}{} & \multicolumn{1}{c}{$M(\mathrm{H\emissiontype{I}})$} \\

\multicolumn{1}{c}{(km s$^{-1}$)} & \multicolumn{1}{c}{} & \multicolumn{1}{c}{($\Mo$)} \\

\hline

30--200 & & $2.1 \times 10^{6}$ \\
50--200 & & $1.4 \times 10^{6}$ \\

\hline

    \end{tabular}
  \end{center}
\end{table}

\newpage

\begin{table}
  \begin{center}
 \caption{Physical properties of molucular loops}\label{tab:loops}
    \begin{tabular}{clllrrrr}

\hline \hline 

\multicolumn{1}{c}{Name} & \multicolumn{1}{c}{Extent in $l$} & \multicolumn{1}{c}{Extent in $b$} & 
\multicolumn{1}{c}{Extent in $V_{\mathrm{LSR}}$} & \multicolumn{1}{c}{$\lambda$\footnotemark[$*$]} & 
\multicolumn{1}{c}{$h$\footnotemark[$*$]} & \multicolumn{1}{c}{$M$} & 
\multicolumn{1}{c}{$\frac{W_{\mathrm{fp}}(\mathrm{\atom{C}{}{12}\atom{O}{}{}})}{W_{\mathrm{tot}}(\mathrm{\atom{C}{}{12}\atom{O}{}{}})}$} \\

\multicolumn{1}{c}{} & \multicolumn{1}{c}{($^{\circ}$)} & \multicolumn{1}{c}{($^{\circ}$)} & 
\multicolumn{1}{c}{(km s$^{-1}$)} & \multicolumn{1}{c}{(pc)} & 
\multicolumn{1}{c}{(pc)} & \multicolumn{1}{c}{($\Mo$)} & 
\multicolumn{1}{c}{} \\

\hline

loop 1 & 356--358 & 0.0--1.5& $-$180--$-$90 & 300 & 220 & 
\multirow{2}{*}{$1.7 \times 10^5$\footnotemark[$\dagger$]} & \multirow{2}{*}{$0.35$\footnotemark[$\dagger$]} \\
loop 2 & 355--356 & 0.5--2.0 & $-$90--$-$40  & 150 & 300 &  & \\
loop 3 & 355--359 & 0.0--2.0 & 30--160  & 600 & 300 & $\sim 3 \times 10^6$ & 0.20 \\

\hline

\multicolumn{4}{@{}l@{}}{\hbox to 0pt{\parbox{85mm}{
  \par\noindent
  Col.(1): Loop name, Col.(2)--(4): Extent in Galactic longitude, latitude and LSR velocity,
  Col.(5): Projected length of the loop,
  Col.(6): Height of the loop from Galactic plane ($b=0^{\circ}$), Col.(7): Mass of the loop,
  Col.(8): \atom{C}{}{12}\atom{O}{}{} integrated intensity ratio of foot point regions to entire region.
  \par\noindent
  \footnotemark[$*$] The distance is assumed to be 8.5 kpc.
  \par\noindent
  \footnotemark[$\dagger$] Loops 1 and 2 are argued together because it is difficult to estimate separately.

}\hss}}

    \end{tabular}
  \end{center}
\end{table}

\newpage

\begin{table}
  \begin{center}
 \caption{Parameters of Miyamoto-Nagai potential modified by Sofue 1996}\label{tab:pot}
    \begin{tabular}{rrrl}

\hline \hline 

\multicolumn{1}{c}{$M_{\mathrm{i}}$} & \multicolumn{1}{c}{$a_{\mathrm{i}}$} & 
\multicolumn{1}{c}{$b_{\mathrm{i}}$} & \multicolumn{1}{c}{component} \\

\multicolumn{1}{c}{($10^{11} \Mo$)} & \multicolumn{1}{c}{(kpc)} & 
\multicolumn{1}{c}{(kpc)} & \multicolumn{1}{c}{} \\

\hline

0.05 & 0.00 & 0.12 & nuclear \\ 
0.10 & 0.00 & 0.75 & bulge   \\ 
1.60 & 6.00 & 0.50 & disk  \\ 
3.00 & 15.00 & 15.00 & halo \\ 

\hline

    \end{tabular}
  \end{center}
\end{table}

\clearpage

\newpage

\newpage

\begin{figure}
  \begin{center}
    \FigureFile(160mm,150mm){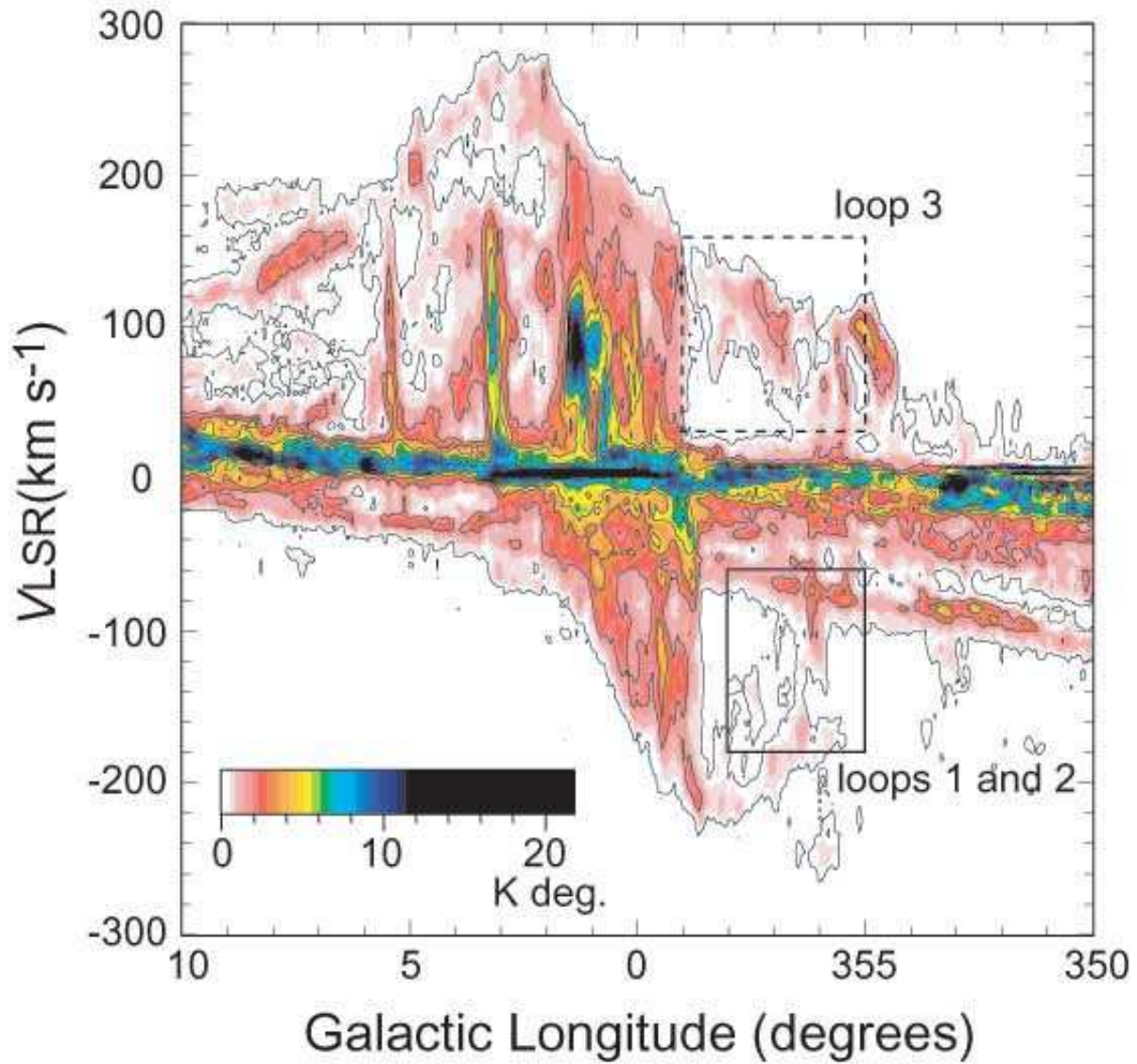}
  \end{center}
  \caption{Galactic longitude-velocity diagram of \atom{C}{}{12}\atom{O}{}{} ($J=1$--$0$) emission 
           observed with NANTEN.
           The integrated galactic latitude range is from $-5\fdg0$ to $5\fdg0$.
           The lowest contour level and the contour intervals are 0.20 K deg and 0.15 K deg, respectively.
           The region of loops 1 and 2 is denoted by the solid line.
           The region of loop 3 is denoted by the broken line.
           (Color Online)}
  \label{GC_lv}
\end{figure}

\newpage

\newpage

\newpage

\newpage

\begin{figure}
  \begin{center}
    \FigureFile(160mm,110mm){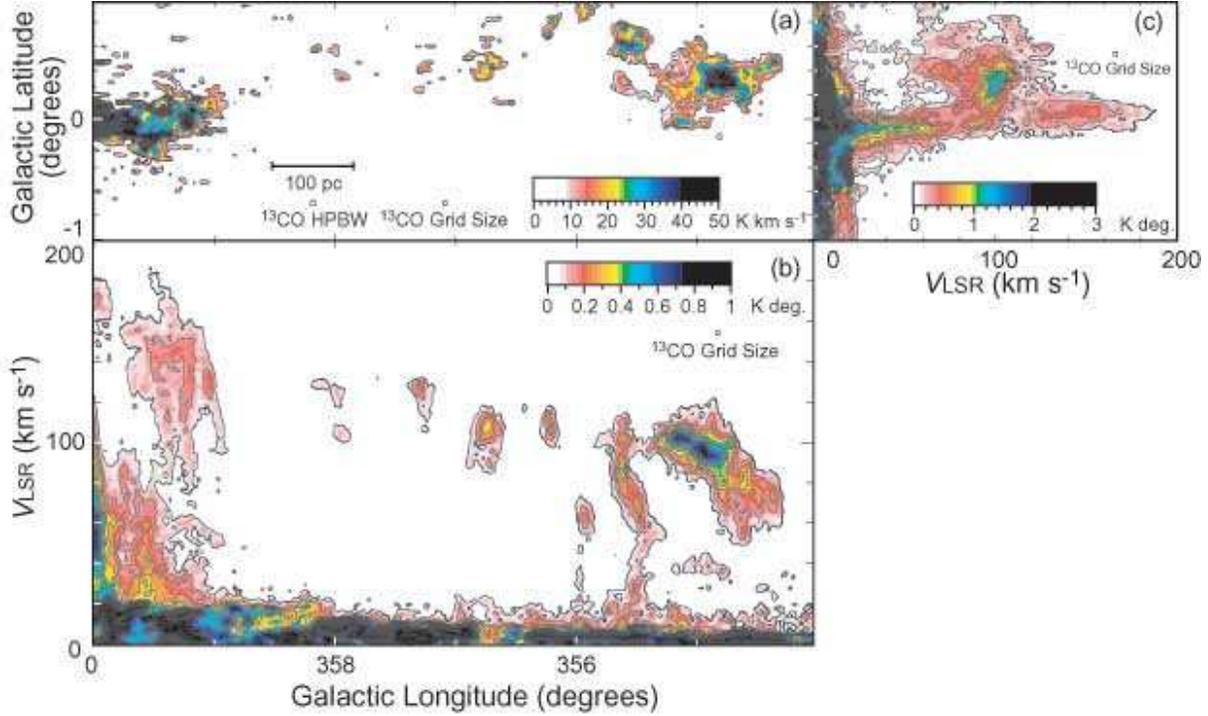}
  \end{center}
  \caption{(a) Velocity-integrated intensity map of \atom{C}{}{13}\atom{O}{}{} ($J=1$--$0$) emission.
           The integrated velocity range is from 30 km s$^{-1}$ to 200 km s$^{-1}$.
           The lowest contour level and the contour intervals are 
           8.0 K km s$^{-1}$ and 10 K km s$^{-1}$, respectively.
           (b) Galactic longitude-velocity diagram of \atom{C}{}{13}\atom{O}{}{} ($J=1$--$0$) emission.
           The integrated galactic latitude range is from $-0\fdg50$ to $1\fdg0$.
           The lowest contour level and the contour intervals are 
           5.9 $\times 10^{-2}$ K deg and 7.9 $\times 10^{-2}$ K deg, respectively.
           (c) Velocity-galactic latitude diagram of \atom{C}{}{13}\atom{O}{}{} ($J=1$--$0$) emission.
           The integrated galactic longitude range is from $354\fdg0$ to $0\fdg0$. 
           The lowest contour level and the contour intervals are 
           8.9 $\times 10^{-2}$ K deg and 0.15 K deg, respectively. 
           (d) Velocity-integrated intensity map of \atom{C}{}{12}\atom{O}{}{} 
           and \atom{C}{}{13}\atom{O}{}{} ($J=1$--$0$) emission.
           The gray scale image and lowest level contour denoted by the thik line indicate \atom{C}{}{12}\atom{O}{}{} emission
           and are same as Figure \ref{12COlbv}a.
           Contours denoted by the thin line indicate \atom{C}{}{13}\atom{O}{}{} emission and 
           are same as Figure \ref{13COlbv}a.
           (e) Galactic longitude-velocity diagram of \atom{C}{}{12}\atom{O}{}{} 
           and \atom{C}{}{13}\atom{O}{}{} ($J=1$--$0$) emission.
           The gray scale image and lowest level contour denoted by the thik line indicate \atom{C}{}{12}\atom{O}{}{} emission
           and are same as Figure \ref{12COlbv}b.
           Contours denoted by the thin line indicate \atom{C}{}{13}\atom{O}{}{} emission and 
           are same as Figure \ref{13COlbv}b.
           (f) Velocity-galactic latitude diagram of \atom{C}{}{12}\atom{O}{}{} 
           and \atom{C}{}{13}\atom{O}{}{} ($J=1$--$0$) emission.
           The gray scale image and lowest level contour denoted by the thik line indicate \atom{C}{}{12}\atom{O}{}{} emission
           and are same as Figure \ref{12COlbv}c.
           Contours denoted by the thin line indicate \atom{C}{}{13}\atom{O}{}{} emission and 
           are same as Figure \ref{13COlbv}c.
           (Color Online)}
  \label{13COlbv}
\end{figure}

\newpage
\begin{figure}
  \begin{center}
    \FigureFile(160mm,125mm){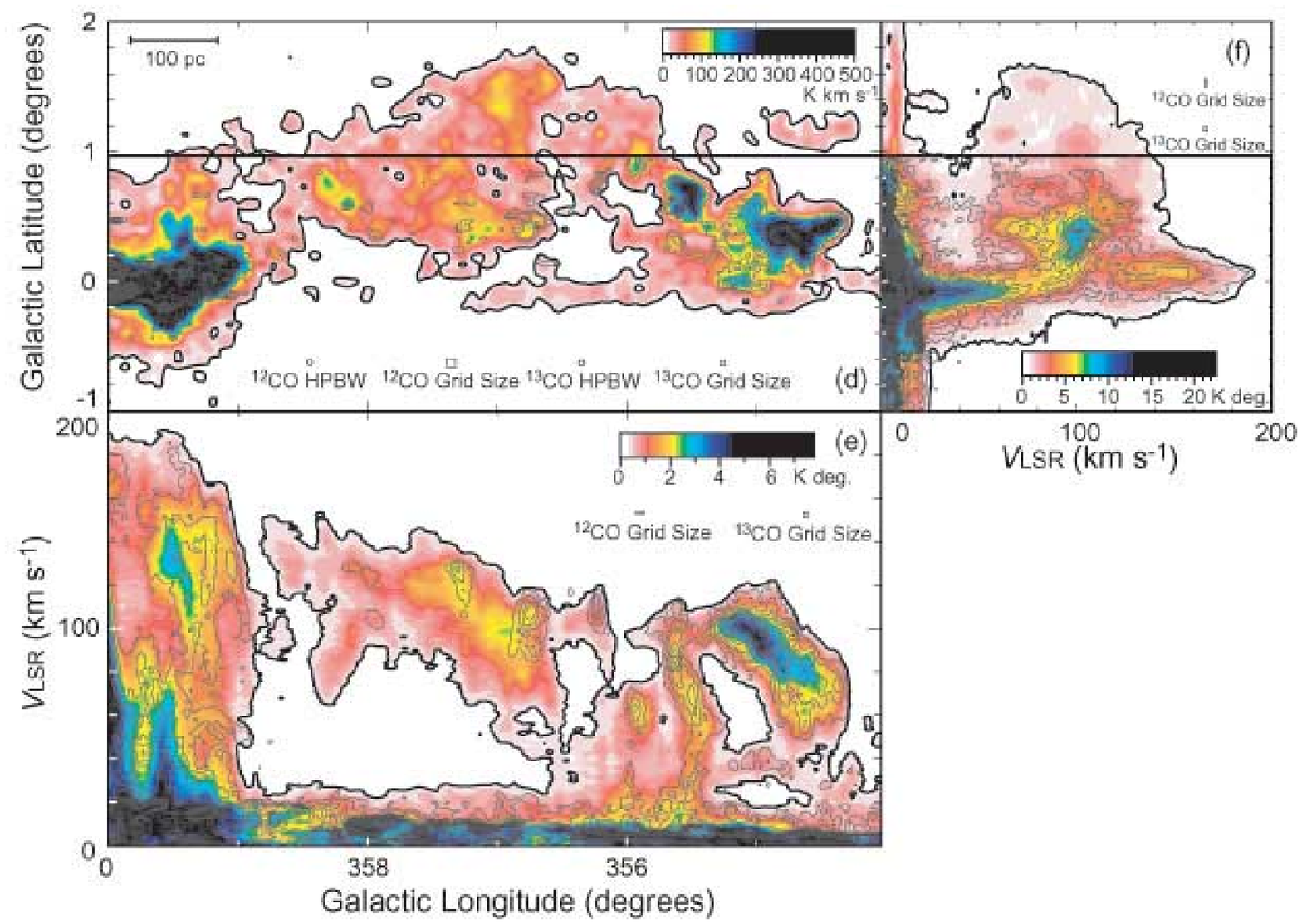}
  \end{center}
  \contcaption{
               (Continued.)
               }
\end{figure}

\newpage

\begin{figure}
  \begin{center}
    \FigureFile(160mm,160mm){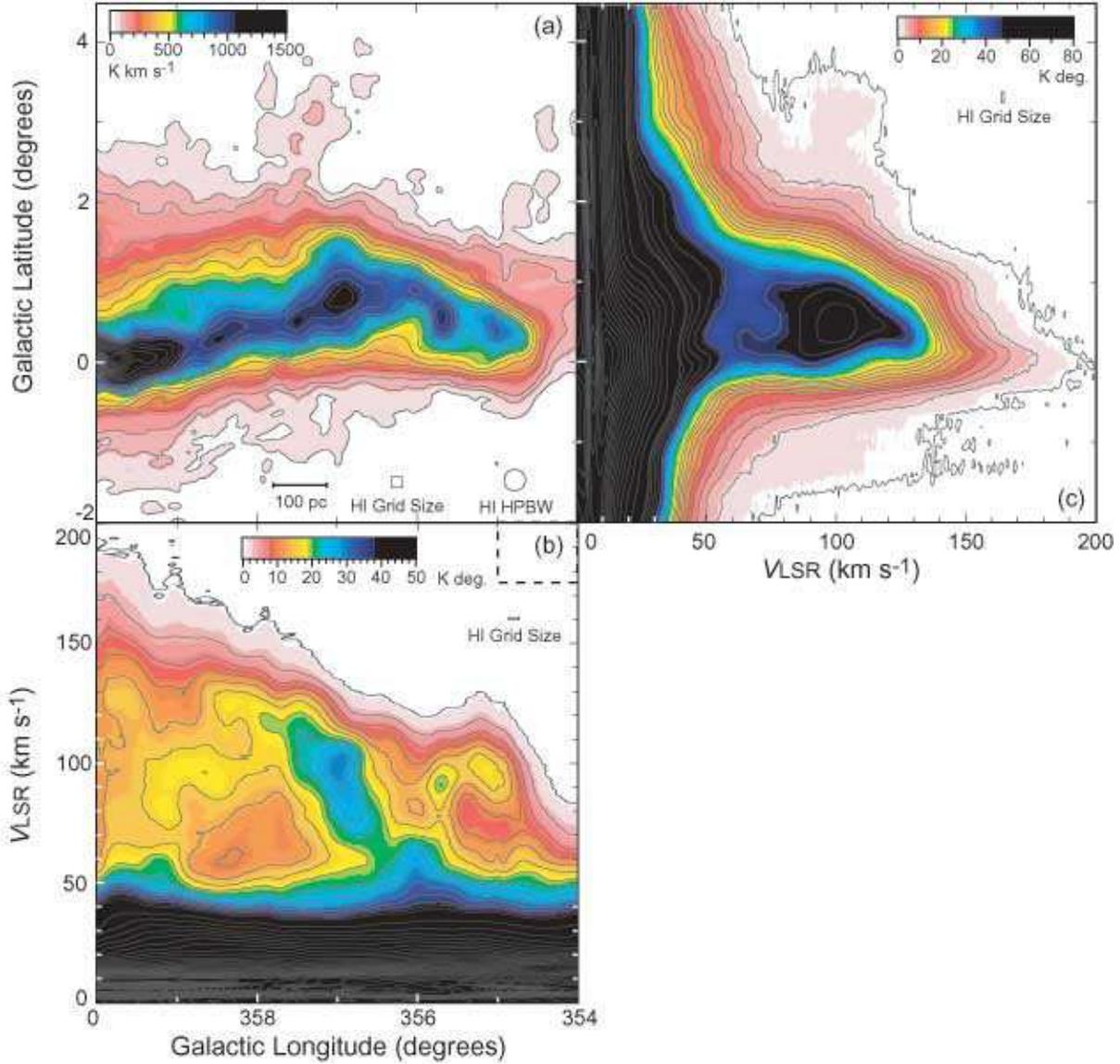}
  \end{center}
  \caption{(a) Velocity-integrated intensity map of H\emissiontype{I} emission. 
           The integrated velocity range is from 50 km s$^{-1}$ to 200 km s$^{-1}$.
           The lowest contour level and the contour intervals are 
           42 K km s$^{-1}$ and 80 K km s$^{-1}$, respectively.
           In addition, 82 K km s$^{-1}$ level is plotted.
           (b) Galactic longitude-velocity diagram of H\emissiontype{I} emission.
           The integrated galactic latitude range is from $-0\fdg50$ to $4\fdg5$.
           The lowest contour level is 0.76 K deg.
           The contour intervals are 3.0 K deg until 43 K deg and then 15 K deg. 
           (c) Velocity-galactic latitude diagram of H\emissiontype{I} emission.
           The integrated galactic longitude range is from $354\fdg0$ to $0\fdg0$
           The lowest contour level is 0.49 K deg.
           The contour intervals are 2.0 K deg until 21 K deg, 8.0 K deg from 21 K deg to 81 K deg
           and then 15 K deg. 
           (Color Online)}
  \label{HIlbv}
\end{figure}

\newpage

\begin{figure}
  \begin{center}
    \FigureFile(160mm,160mm){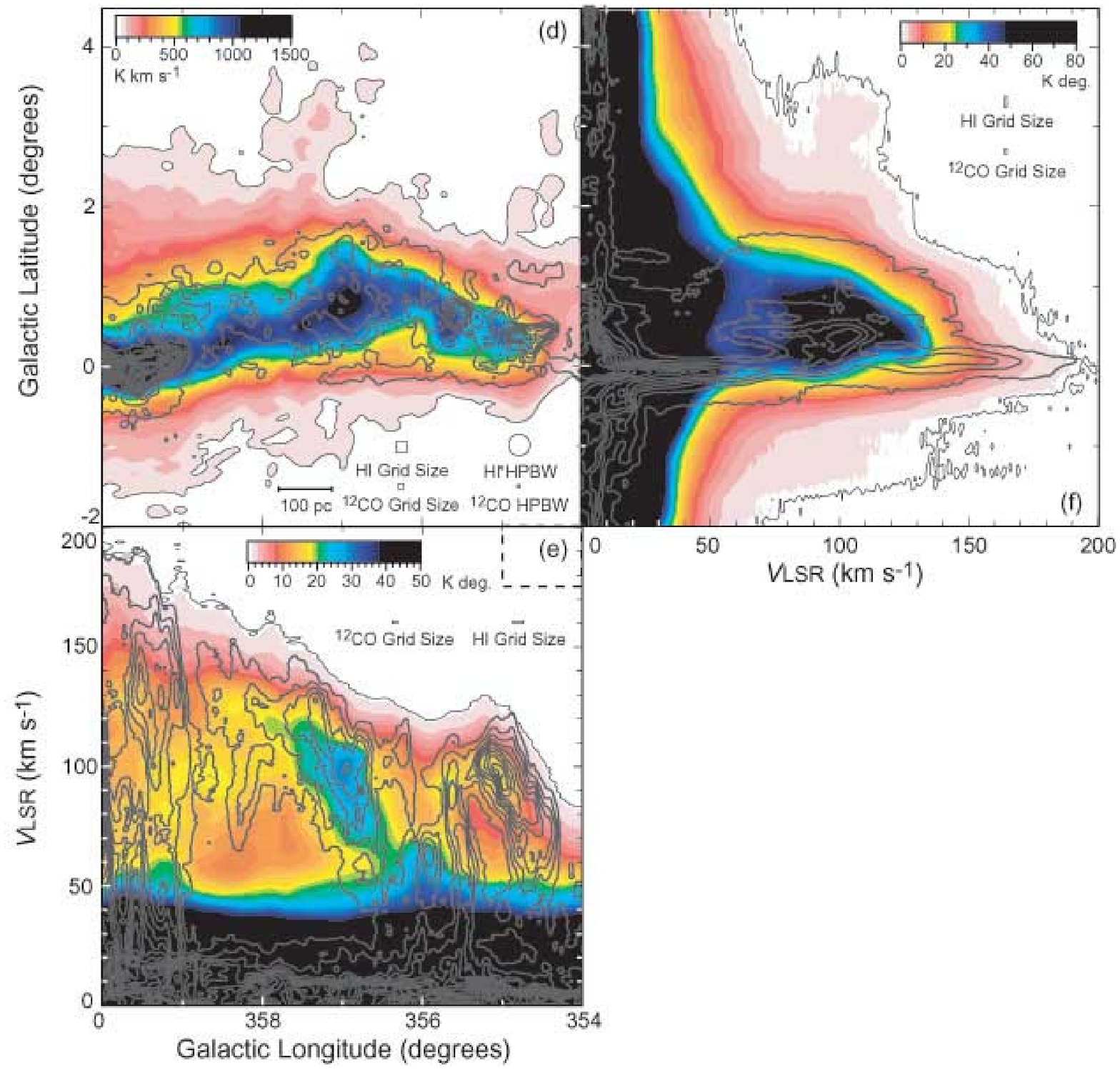}
  \end{center}
  \contcaption{(d) Velocity-integrated intensity map of H\emissiontype{I} and \atom{C}{}{12}\atom{O}{}{} ($J=1$--$0$) emission.
               The gray scale image and lowest level contour denoted by the black line indicate H\emissiontype{I} emission 
               and are same as Figure \ref{HIlbv}a.
               Contours denoted by the gray line indicate \atom{C}{}{12}\atom{O}{}{} emission.
               The integrated velocity range is from 30 km s$^{-1}$ to 200 km s$^{-1}$.
               The lowest contour level and the contour intervals are 
               10 K km s$^{-1}$ and 60 K km s$^{-1}$, respectively.
               (e) Galactic longitude-velocity diagram of H\emissiontype{I} and \atom{C}{}{12}\atom{O}{}{} ($J=1$--$0$) emission.
               The gray scale image and lowest level contour denoted by the black line indicate H\emissiontype{I} emission 
               and are same as Figure \ref{HIlbv}b.
               Contours denoted by the gray line indicate \atom{C}{}{12}\atom{O}{}{} emission 
               and are same as Figure \ref{HIlbv}b.
               (f) Velocity-galactic latitude diagram of H\emissiontype{I} and \atom{C}{}{12}\atom{O}{}{} ($J=1$--$0$) emission.
               The gray scale image and lowest level contour denoted by the black line indicate H\emissiontype{I} emission 
               and are same as Figure \ref{HIlbv}c.
               Contours denoted by the gray line indicate \atom{C}{}{12}\atom{O}{}{} emission 
               and are same as Figure \ref{HIlbv}c.
               (Color Online)
               }
\end{figure}

\newpage

\newpage

\newpage

\end{document}